# Delta-$T$ noise for fractional quantum Hall states at different filling factor

Giacomo Rebora,[1,2,*] Jérôme Rech,[3] Dario Ferraro,[1,2] Thibaut Jonckheere,[3] Thierry Martin,[3] and Maura Sassetti[1,2]
[1]*Dipartimento di Fisica, Università di Genova, Via Dodecaneso 33, 16146, Genova, Italy*
[2]*CNR-SPIN, Via Dodecaneso 33, 16146, Genova, Italy*
[3]*Aix Marseille Univ, Université de Toulon, CNRS, CPT, IphU, AMUtech, Marseille 13288, France*

The current fluctuations due to a temperature bias, i.e., the delta-$T$ noise, allow one to access properties of strongly interacting systems which cannot be addressed by the usual voltage-induced noise. In this work, we study the full delta-$T$ noise between two different fractional quantum Hall edge states, with filling factors ($\nu_L$, $\nu_R$) in the Laughlin sequence, coupled through a quantum point contact and connected to two reservoirs at different temperatures. We are able to solve exactly the problem for all couplings and for any set of temperatures in the specific case of a hybrid junction (1/3, 1). Moreover, we derive a universal analytical expression which connects the delta-$T$ noise to the equilibrium one valid for all generic pairs ($\nu_L$, $\nu_R$) up to the first order in the temperature mismatch. We expect that this linear term can be accessible in today's experimental setups. We describe the two opposite-coupling regimes focusing on the strong one, which corresponds to a nontrivial situation. Our analysis on delta-$T$ noise allows us to better understand the transport properties of strongly interacting systems and to move toward more involved investigations concerning the statistics and scaling dimension of their emergent excitations.

## I. INTRODUCTION

The fractional quantum Hall (FQH) effect takes its roots in the strong correlations among electrons due to the Coulomb interaction [1]. When a two-dimensional electron gas (2DEG) is subjected to an intense perpendicular magnetic field at low temperatures, electrons can form a quantum state whose excitations are described by fractional quantum numbers, including fractional charge [2] and fractional statistics [3,4]. Current fluctuation measurements have been crucial to demonstrate the existence of these peculiar fractionally charged excitations [5–8] and to access their nontrivial statistics [9–12].

On a general ground, noise measurements are fundamental tools for the understanding of mesoscopic devices. When the considered system is in equilibrium, i.e., when no voltage bias is applied, the average current flowing across it is zero. However, at finite temperature a contribution to the noise, referred to as the thermal or Johnson-Nyquist noise, is always present [13,14] due to the thermal motion of carriers. Conversely, when the voltage bias applied to the mesoscopic conductor is dominant with respect to the temperature, the current fluctuations can be described in terms of the nonequilibrium shot noise [15]. It originates from the discrete nature of the electric charge and it has been used extensively to investigate electron correlations in quantum liquids.

In recent years, systems connected to reservoirs kept at different temperatures, when no voltage is applied, have been experimentally [16–21] and theoretically [22–29] considered. The presence of a temperature gradient leads to a nonequilibrium contribution to the charge current noise known as delta-$T$ noise. It is expected to carry additional information on quantum correlations inside the systems since it allows us to address directly the tunneling density of states.

In this context, the delta-$T$ noise was recently studied for FQH systems in a quantum point contact (QPC) geometry. Here, the delta-$T$ contribution to the noise is quadratic, due to the symmetry of the considered setup under the exchange of reservoirs, and it was found that the tunneling of quasiparticles is associated with negative values of the delta-$T$ noise [23]. In other words, the nonequilibrium noise induced by the temperature mismatch between the two channels of this correlated state turns out to be smaller than the equilibrium noise. Conversely, when electrons tunnel from one edge to the other, the delta-$T$ noise becomes positive. While it was shown that interactions alone could not account for such negative delta-$T$ noise, it was put forward that, in several recent works, negative contributions to the noise were attributed to braiding effects, albeit in different configurations [30,31]. Although a direct connection could not be formally obtained at the time, it was deemed plausible that such negative delta-$T$ noise could arise as an effect of the anyonic statistics of the particles tunneling through.

More recently, this connection was further explored and shown to be merely a byproduct of the true physical mechanism at play [29]. There, the authors carried out a detailed

---

investigation of the delta-$T$ noise in weakly coupled identical one-dimensional chiral interacting systems, covering fractional quantum Hall as well as quantum spin Hall edge states. It was then shown that the sign of the delta-$T$ noise was uniquely determined by the scaling dimension of the tunneling operator $\Delta_T$. It follows that a device where the tunneling is dominated by a process with scaling dimension $\Delta_T < 1/2$ would yield a negative delta-$T$ noise, irrespective of the nature of the tunneling excitations or their quantum statistics. As it turns out, however, for fractional quantum Hall edge states, tunneling processes with a scaling dimension $\Delta_T < 1/2$ involve quasiparticles with a scaling dimension $\Delta < 1/4$ (assuming a symmetric junction), whose statistical angle is then bounded as $|\theta| \leqslant \pi/2$. It follows that in such devices negative delta-$T$ noise is observed for "boson-like" quasiparticles. This connection between delta-$T$ noise and scaling dimension of the tunneling operator was first hinted at in Ref. [27] before being formally demonstrated in Ref. [29]. It does make sense that such a nonequilibrium thermal noise contribution grants access to properties of the energy distribution of the tunneling events, which is itself encoded in the scaling dimension of the tunneling operator.

The study of the delta-$T$ noise can thus be exploited in order to investigate the properties of strongly interacting systems, in a way that cannot be addressed by the usual voltage-induced noise. Moreover, the absence of any bias voltage allows one to discard some of the nonuniversal effects (changes in the electrostatic properties of the point contact, slight modifications of the edge confining potential) that typically make the comparison with experimental data all the more difficult.

In this work, we study the delta-$T$ noise generated by a temperature gradient between two different FQH edge states in the Laughlin regime, whose filling factors are indicated in the following with the short notation $(\nu_L, \nu_R)$. Differently from the analysis carried out so far, which has usually been focused on some perturbative expansion, we do not assume *a priori* a specific strength of the coupling between the two systems. This approach allows us to solve exactly the problem for all couplings and for any set of temperatures in a hybrid junction with specific filling factors $(1/3, 1)$. This configuration has recently been the focus of great attention due to the observation of exotic Andreev reflection processes involving fractional quasiparticles [32]. From the theoretical point of view, the relevance of this case relies on the fact that it can be exactly solved through refermionization [33,34]. According to this, the delta-$T$ noise can be obtained numerically for any set of temperatures of the two Hall bars and for all coupling strengths. Moreover, our results suggest that, unlike the homogeneous case $\nu_L = \nu_R$, linear-in-$\Delta T$ contributions dominate here, at low temperature bias.

This motivated us to focus on the situation where a small mismatch in the temperatures between the two Hall bars is considered, which turns out to also be the most relevant for possible experimental implementations. This then allows us to focus on the first (linear) order in the temperature expansion, but taking into account all orders in the tunneling amplitudes. As it turns out, this analytical treatment can be extended to all filling factor pairs $(\nu_L, \nu_R)$ and leads to a universal relation which connects the delta-$T$ noise to the equilibrium noise. As examples of the validity of the present approach, we focus on two opposite coupling regimes. In the weak-coupling regime, the two edges are almost decoupled and a current impinging upon the QPC is almost perfectly reflected. The opposite regime of strong coupling is more complex and interesting because it corresponds to a nontrivial situation, where perfect transmission of the current is reduced by the scattering of fractional quasiparticles resulting from the coupling of the two systems. This opposite regime requires a weak-strong duality relation [35] in order to be addressed.

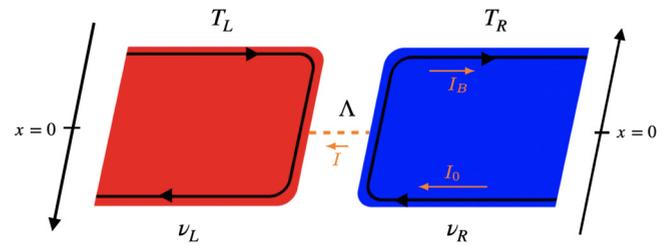

FIG. 1. Sketch of two fractional quantum Hall systems with different filling factors ($\nu_L$ and $\nu_R$) belonging to Laughlin sequence and at different temperatures ($T_L$ and $T_R$). Counterpropagating edge states are coupled via a point-like tunneling (orange dashed line) placed in $x = 0$ with $\Lambda$ being the constant tunneling amplitude strength. Notice that the $x$ axis follows the chirality of the channels. In orange, we have highlighted the impinging current on the QPC $I_0$, related to the quantized Hall conductance, the tunneling current $I$ and the backscattering one $I_B$.

This analysis aims to deepen our knowledge on delta-$T$ noise in order to move towards more involved applications concerning the possibility of investigating the statistics [36,37] and the scaling dimension of the carriers [27,29].

This paper is organized as follows: In Sec. II, we introduce the model for the hybrid FQH junction where the counterpropagating edge states are coupled by a local tunneling region, as depicted in Fig. 1. This description sets the basis for the exploration of the weak-coupling regime. In Sec. III we write down an explicit expression for the current fluctuations in the presence of a temperature gradient due to different temperatures, namely, the delta-$T$ noise. To investigate the strong-coupling regime we exploit the duality relation in Sec. IV. Then, our purpose in Sec. V is to study what happens for the delta-$T$ noise in a particular QH junction with ($\nu_L = 1/3$, $\nu_R = 1$) for all coupling strengths and for any temperature. To be closer to experiments and to consider a more generic hybrid system $(\nu_L, \nu_R)$ in Sec. VI we give a universal expression for the delta-$T$ up to the first order in temperature gradient, for all tunneling regimes. Then, in Sec. VII, we explore the first order delta-$T$ noise in the two opposite tunneling regimes by focusing, first, on an asymmetric junction with fixed filling factors $(1/3, 1)$ and, second, for general values of $\nu_L$ and $\nu_R$. Finally, in Sec. VIII we summarize our results. Analytical details of the derivation of the obtained results are reported in the Appendixes.

## II. MODEL FOR THE JUNCTION

We consider two FQH bars at different filling factors $\nu_\alpha$ ($\alpha = L, R$) belonging to the Laughlin sequence, i.e.,

$\nu_\alpha = 1/(2n + 1)$ ($n \in \mathbb{N}$) [1,2,38]. They are kept at two different temperatures $T_L$ and $T_R$ and coupled through a point-like tunneling region, as depicted in Fig. 1. The edge states of such a system are described in terms of a hydrodynamical model [39] by a chiral Luttinger liquid free Hamiltonian of the form ($\hbar = k_B = 1$)

$$H^{(0)} = H_L^{(0)} + H_R^{(0)} = \sum_{\alpha=L,R} \frac{v_\alpha}{4\pi} \int dx [\partial_x \phi_\alpha(x)]^2, \quad (1)$$

where $\phi_\alpha$ are the bosonic fields describing the counterpropagating modes traveling along the edge of the left and right QH bars. They satisfy the usual commutation relation $[\phi_\alpha(x), \phi_\beta(y)] = i\pi \delta_{\alpha\beta} \text{sgn}(x - y)$ with $\alpha, \beta = L, R$ [39].

Quite generally, the velocities $v_L$ and $v_R$ along the two edges can be different. However, in what follows, we focus on the situation where the tunneling occurs at a specific point, allowing us to rescale the position coordinates independently for the two bosonic fields. This in turn enables us to alter the velocities at will, so that, for sake of simplicity, we assume the same propagation velocity for the two edges ($v_L = v_R = v$). The chiral bosonic particle-hole collective excitations described by the fields $\phi_\alpha$ ($\alpha = L, R$) in Eq. (1) are related to the particle density of the channel $\rho_\alpha$ through the relation

$$\rho_\alpha(x) = \frac{\sqrt{\nu_\alpha}}{2\pi} \partial_x \phi_\alpha(x). \quad (2)$$

Using the conventional bosonization technique, the electron annihilation operator $\psi_\alpha(x)$ can be expressed in terms of $\phi_\alpha(x)$ as [40,41]

$$\psi_\alpha(x) = \frac{\mathcal{F}_\alpha}{\sqrt{2\pi a}} e^{-i\frac{1}{\sqrt{\nu_\alpha}}\phi_\alpha(x)}, \quad (3)$$

with $a$ a short-distance cutoff and $\mathcal{F}_\alpha$ being the Klein factor [42]. In the following we omit this latter factor since it plays no role in the calculation of the correlators [43].

We assume that the two QH systems are coupled via a quantum point contact (QPC) placed at position $x = 0$, which allows local tunneling between the two counterpropagating edges. In practice, there is no bulk Hall fluid in between the two edge states, so that the only allowed tunneling process involves electrons. This configuration, where only electrons can locally tunnel from one lead to the other, is described by the tunneling Hamiltonian

$$H_\Lambda = \Lambda \int dx \delta(x) \psi_R^\dagger(x) \psi_L(x) + \text{H.c.}$$
$$= \frac{\Lambda}{2\pi a} e^{i\frac{1}{\sqrt{\nu_R}}\phi_R(0)} e^{-i\frac{1}{\sqrt{\nu_L}}\phi_L(0)} + \text{H.c.}, \quad (4)$$

where the second line stands from Eq. (3) and $\Lambda$ is a constant tunneling amplitude strength [44–47]. Notice that the QPC description conventionally used to model the tunneling between QH edge states is valid as long as the width of the tunneling region is of the order of the magnetic length. This situation is typically quite well achieved in experiments devoted to noise measurement in quantum Hall systems [7]. Generalization of this picture towards extended contacts have been considered [48,49]. In this situation, which is out of the aim of the present work, additional effects such as disorder at the level of the contact or interferences due to the formation of Aharonov-Bohm loops could come into play and need to be properly taken into account.

The current operator $I(t)$ describing the tunneling current flowing from one edge state to the other is obtained from the tunneling Hamiltonian and reads

$$I(t) = -e\dot{N}_R = ie[N_R, H_\Lambda]$$
$$= ie\Lambda \psi_R^\dagger(t)\psi_L(t) + \text{H.c.}$$
$$= ie\frac{\Lambda}{2\pi a} e^{i\frac{1}{\sqrt{\nu_R}}\phi_R(t)} e^{-i\frac{1}{\sqrt{\nu_L}}\phi_L(t)} + \text{H.c.}, \quad (5)$$

with $N_\alpha = \int dx \rho_\alpha(x)$ being the particle numbers on each edge and where the notation only keeps track of the time dependence of the field operators omitting the fact that they are evaluated at the QPC in $x = 0$. Notice that, in the second line, we have rewritten the fermionic operators in terms of the bosonic ones following the prescription in Eq. (3). The current operator can be rewritten as

$$I(t) = -e\dot{N}_R = -e \int_{-\infty}^{+\infty} dx \partial_t \rho_R(x, t)$$
$$= ie \int_{-\infty}^{+\infty} dx [\rho_R(x, t), H_\Lambda]. \quad (6)$$

Since $H_\Lambda$ only involves fields at the position of the QPC, the resulting commutator is nonzero only close to $x = 0$. This allows us to write

$$I(t) = ie \int_{0^-}^{0^+} dx [\rho_R(x, t), H_\Lambda]$$
$$= -e \int_{0^-}^{0^+} dx \partial_t \rho_R(x, t). \quad (7)$$

Exploiting the chirality and linear dispersion of edge states, one can readily write $\partial_t \rho_R(x, t) = -v \partial_x \rho_R(x, t)$ so that the current can be expressed in terms of the densities right before ($x = 0^-$) and right after ($x = 0^+$) the QPC as

$$I(t) = ev[\rho_R(0^+, t) - \rho_R(0^-, t)]. \quad (8)$$

We now consider a suitable rotation in the field space [36],

$$\begin{pmatrix} \varphi_L(x) \\ \varphi_R(x) \end{pmatrix} = \begin{pmatrix} \cos\theta & \sin\theta \\ -\sin\theta & \cos\theta \end{pmatrix} \begin{pmatrix} \phi_L(x) \\ \phi_R(x) \end{pmatrix}, \quad (9)$$

with angle satisfying

$$\sin 2\theta = \frac{v_R - v_L}{v_R + v_L}. \quad (10)$$

The free Hamiltonian and the coupling term are then rewritten as

$$H = H^{(0)} + H_\Lambda = \sum_{\alpha=L,R} \frac{v}{4\pi} \int dx [\partial_x \varphi_\alpha(x)]^2$$
$$+ \frac{\Lambda}{2\pi a} e^{i\frac{1}{\sqrt{g}}[\varphi_R(0) - \varphi_L(0)]} + \text{H.c.}, \quad (11)$$

which corresponds to the tunneling between two identical chiral Luttinger liquids with effective filling factor

$$g^{-1} = \frac{1}{2}\left(\frac{1}{\nu_L} + \frac{1}{\nu_R}\right). \quad (12)$$

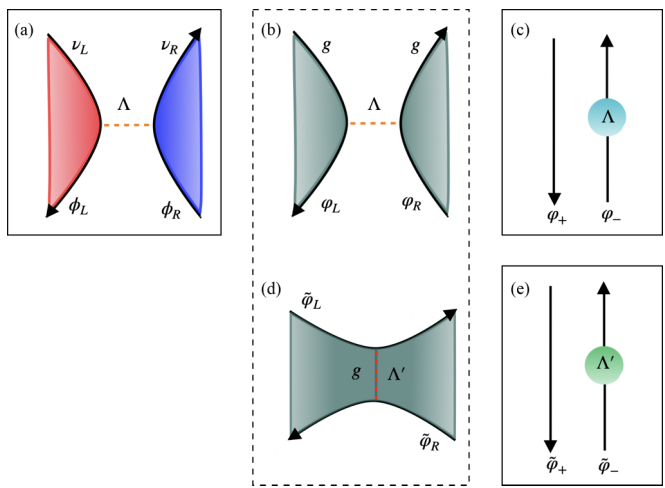

FIG. 2. (a) Scheme of the two FQH systems with filling factors $\nu_L$ and $\nu_R$ whose edges are described by the bosonic fields $\phi_L$ and $\phi_R$. (b) This panel is obtained through the rotation in Eq. (9) which maps the original junction, with tunneling parameter $\Lambda$, between two different FQH liquids to one between FQH with the same effective filling factor $g$. (c) The rotation in Eq. (13) decouples the problem into two separate ones: one free channel and the other describing a channel with the QPC referred as a localized impurity. (d) The duality relation in Eq. (21) allows us to map the problem of two FQH liquids in the weak-coupling regime, where electrons can tunnel, to the strong coupling one described by a single FQH liquid, with filling factor $g$ and tunneling parameter $\Lambda'$. In this configuration, the quasiparticles provide the dominant contribution to the tunneling. (e) The further rotation in Eq. (24) decouples the problem into two separate ones in analogy with panel (c).

In this way we map the problem of electron tunneling between different FQH edges to the problem of electron tunneling between two identical chiral Luttinger liquids with the same effective filling factor $g$. This mapping can be visualized passing from Fig. 2(a) to Fig. 2(b).

Moreover, a further transformation is introduced:

$$\varphi_+(x) = \frac{1}{\sqrt{2}}[\varphi_L(x) + \varphi_R(x)],$$

$$\varphi_-(x) = \frac{1}{\sqrt{2}}[\varphi_L(x) - \varphi_R(x)]. \quad (13)$$

According to this, the free Hamiltonians become

$$H_{L/R}^{(0)} = \frac{v}{4\pi} \int dx \bigg\{ \frac{1}{2}(\cos\theta \mp \sin\theta)^2 [\partial_x \varphi_+(x)]^2$$

$$+ \frac{1}{2}(\sin\theta \pm \cos\theta)^2 [\partial_x \varphi_-(x)]^2$$

$$\pm (\cos^2\theta - \sin^2\theta)[\partial_x \varphi_+(x)][\partial_x \varphi_-(x)] \bigg\}, \quad (14)$$

and their sum depends on the new fields $\varphi_\pm$ separately,

$$H_L^{(0)} + H_R^{(0)} = \frac{v}{4\pi} \int dx \{[\partial_x \varphi_+(x)]^2 + [\partial_x \varphi_-(x)]^2\}. \quad (15)$$

Then, the total Hamiltonian of Eq. (11) is rewritten as

$$H = H_+ + H_-$$

$$= \frac{v}{4\pi} \int dx [\partial_x \varphi_+(x)]^2 + \frac{v}{4\pi} \int dx [\partial_x \varphi_-(x)]^2$$

$$+ \left[ \frac{\Lambda}{2\pi a} e^{i\sqrt{\frac{2}{g}}\varphi_-(0)} + \text{H.c.} \right]. \quad (16)$$

The transformation of Eq. (13) allows us to decouple the problem of the two FQH liquids with same $g$ into two separate ones [see Fig. 2(c)]: the first depending only on the free field $\varphi_+$ and the second one which includes the tunneling contribution and that can be written only in terms of the field $\varphi_-$.

From a physical point of view, by looking at Fig. 2(a), when the two systems are totally decoupled (i.e., $\Lambda = 0$) and if an electron is sent from one of the two QH bars, it is perfectly reflected at the contact and there is no net current flowing through the junction. However, if we consider a weak-coupling limit, for which the two edges are almost decoupled (i.e., $\Lambda$ is small but nonzero), the electrons are allowed us to jump from one side to other.

## III. DELTA-$T$ NOISE

The expectation value of the current operator in Eq. (5) is given by ($k_B = 1$)

$$\mathcal{I} = \frac{1}{Z} \text{Tr} \left\{ \exp\left[ -\sum_{\alpha=L,R} \frac{H_\alpha^{(0)}}{T_\alpha} \right] I(t) \right\}, \quad (17)$$

with

$$Z = \text{Tr} \left\{ \exp\left[ -\sum_{\alpha=L,R} \frac{H_\alpha^{(0)}}{T_\alpha} \right] \right\}. \quad (18)$$

Here, we are assuming that, at the time $t \to -\infty$, the tunneling is switched off and the two bars are at thermal equilibrium. This leads to the initial density matrix $\varrho_0 = (1/Z)\exp[-\sum_\alpha H_\alpha^{(0)}/T_\alpha]$ [50]. The tunneling is then turned on, establishing a stationary current. Moreover, since we are considering no voltage bias and local tunneling, the total net current $\mathcal{I}$ is zero independently of the respective temperatures of the two edges. This is due to the fact that the transmission function is energy independent and electrons and holes contribute equally [51,52]. However, since the finite temperature always leads to a nonzero contribution to the noise, through thermal fluctuations [13,14], the current-current fluctuations do not vanish. This charge current noise induced by the mismatch in the temperatures has been dubbed delta-$T$ noise. It has a purely thermal origin, but it is only generated in nonequilibrium situations [17–19,23].

The zero-frequency current noise can be written as

$$S(T_L, T_R)$$

$$= 2 \int_{-\infty}^{+\infty} d\tau \langle \Delta I(\tau) \Delta I(0) \rangle$$

$$= 2 \int_{-\infty}^{+\infty} d\tau \left[ \frac{1}{Z} \text{Tr} \left\{ \exp\left[ -\sum_{\alpha=L,R} \frac{H_\alpha^{(0)}}{T_\alpha} \right] \Delta I(\tau) \Delta I(0) \right\} \right], \quad (19)$$

where $\Delta I(t) = I(t) - \mathcal{I}$.

In the following, we also consider the temperature parametrization

$$T_R = T, \quad T_L = T_R + \Delta T, \qquad (20)$$

which is convenient from an experimental point of view. Here, the temperature of one of the two FQH sample is kept fixed at $T$, while the other can be varied with $\Delta T$ either positive or negative.

## IV. DUALITY RELATION

Let us focus now on the configuration of Fig. 2(b), after the change of basis introduced in Eq. (9), where we move from an inhomogeneous QH junction to two separate FQH liquids at the same filling factor $g$. In this effective picture, as the tunneling amplitude increases up to the limit value $\Lambda \to \infty$ (i.e., strong coupling), we switch from the two identical, but separate, FQH liquids to a unique one [see Fig. 2(d)]. In analogy to what happens for the Laughlin states, the dynamics of the point contact evolves from being dominated by electron tunneling at weak coupling, where the fluid is pinched off, to a strong-coupling regime where quasiparticle tunneling dominates [7,53]. This process is embodied by a powerful electron-quasiparticle duality [44,54] which reflects the duality relation between the weak- and strong-coupling limits. In particular, the strong-coupling limit is accessible through a weak-strong duality transformation [35,55,56]. This fact is graphically described in Fig. 2 by the central panels surrounded by a dashed line. In this limit, the fields $\varphi_L$ and $\varphi_R$ can be written in terms of dual fields $\widetilde{\varphi}_L$ and $\widetilde{\varphi}_R$ defined as

$$\varphi_L(x) = \widetilde{\varphi}_L(x)\Theta(-x) + \widetilde{\varphi}_R(x)\Theta(x),$$
$$\varphi_R(x) = \widetilde{\varphi}_L(x)\Theta(x) + \widetilde{\varphi}_R(x)\Theta(-x), \qquad (21)$$

with $\Theta(\pm x)$ being the step function. This dual transformation can be geometrically understood by thinking about the fact that, in the strong limit, the previous bosonic states $\varphi_L$ and $\varphi_R$ are mixed since now there is only one QH liquid. This nonlocal relation, due to the step function, recall the starting point of two different QH sample separated by a QPC.

Due to the above considerations the total Hamiltonian describing the dynamics and the coupling of these fields is now

$$\widetilde{H} = \sum_{\alpha=L,R} \frac{v}{4\pi} \int dx [\partial_x \widetilde{\varphi}_\alpha(x)]^2$$
$$+ \frac{\Lambda'}{2\pi a} e^{i\sqrt{g}[\widetilde{\varphi}_R(0) - \widetilde{\varphi}_L(0)]} + \mathrm{H.c.} \qquad (22)$$

Notice that this last Hamiltonian is the dual of the one in Eq. (11) where we have considered the substitution $g \to 1/g$ due to the electron-quasiparticle mapping and we have introduced an effective coupling parameter $\Lambda'$. However, the two tunneling strengths $\Lambda$ and $\Lambda'$ are not independent as they are connected by the relation [57]

$$\left(\frac{\Lambda'}{\omega_c a}\right) = \left[2^{-2g+1}\Gamma^g\left(1+\frac{1}{g}\right)\Gamma(1+g)\right]\left(\frac{\Lambda}{\omega_c a}\right)^{-g}, \quad (23)$$

where $\omega_c = v/a$ is a high-energy cutoff and $\Gamma(x)$ is the Euler Gamma function of a given argument $x$. Due to the inverse proportionality between $\Lambda$ and $\Lambda'$, focusing on the $\Lambda \to \infty$ limit is equivalent to consider $\Lambda' \to 0$ and vice versa, consistently with the discussed weak-strong-coupling duality.

The formulation of the problem in terms of the dual fields $\widetilde{\varphi}_{L/R}$ in the strong-coupling limit ($\Lambda \to \infty$ or $\Lambda' \to 0$) has the advantage that these are now free fields. The quasiparticles which tunnels are noninteracting and carry a charge $e^* = ge$. Quite remarkably, these effective fractionally charged excitations correspond neither to electrons nor to quasiparticles of the isolated Hall fluids but instead to complicated nonlocal objects emerging from the dynamics of the two strongly coupled edge channels as a whole.

By introducing the fields

$$\widetilde{\varphi}_\pm(x) = \frac{1}{\sqrt{2}}[\widetilde{\varphi}_L(x) \pm \widetilde{\varphi}_R(x)], \qquad (24)$$

the Hamiltonian of Eq. (22) leads again to two decoupled systems [see Fig. 2(e)]

$$\widetilde{H} = \widetilde{H}_+ + \widetilde{H}_-, \qquad (25)$$

where

$$\widetilde{H}_+ = \frac{v}{4\pi}\int dx[\partial_x \widetilde{\varphi}_+(x)]^2,$$
$$\widetilde{H}_- = \frac{v}{4\pi}\int dx[\partial_x \widetilde{\varphi}_-(x)]^2 + \left[\frac{\Lambda'}{2\pi a}e^{i\sqrt{2g}\widetilde{\varphi}_-(0)} + \mathrm{H.c.}\right]. \quad (26)$$

## V. EXACT SOLUTION FOR TUNNELING IN A (1/3, 1) JUNCTION

In this section we focus on a junction between a normal metal ($\nu_R = 1$) and a FQH state with filling factor $\nu_L = 1/3$ in the presence of a temperature difference between the Hall bars. This case can be exactly solved via refermionization for the entire range of couplings and temperatures, allowing us to evaluate the delta-$T$ noise exactly.

### A. Refermionization

The case $\nu_R = 1$ and $\nu_L = 1/3$ leads to a description, in terms of an effective filling factor $g = 1/2$ [see Eq. (12)]. In the rotated basis, and taking into account the duality, one can consider a tunneling Hamiltonian proportional to the factor $e^{i\widetilde{\varphi}_-}$ [see Eq. (26)], which looks like an electronic operator. It thus becomes possible to introduce a new fermionic field and reexpress the tunneling term accordingly, ultimately allowing us to diagonalize exactly the Hamiltonian, and therefore account for tunneling at all orders.

This idea of refermionization was first introduced in Ref. [33] in the framework of quantum dissipative systems, and subsequently applied to the case FQH states [34]. It amounts to refermionizing the bosonic field $\widetilde{\varphi}_-$ so that the tunneling term of the Hamiltonian $\widetilde{H}_-$ in Eq. (26) now takes the form

$$\frac{\Lambda'}{2\pi a}e^{i\widetilde{\varphi}_-(0)} + \mathrm{H.c.} \longrightarrow \frac{\Lambda'}{\sqrt{2\pi a}}f\psi^\dagger(0) + \mathrm{H.c.}, \qquad (27)$$

where we remind the reader that $x = 0$ is the position of the QPC. Here $f$ is an auxiliary (Majorana) fermion field, introduced in the same spirit as Klein factors in Eq. (3), which arises from the proper handling of the zero modes of the bosonic fields. In particular, the new fields $\psi(x)$ and $f$ satisfy the following set of equations of motion:

$$-i\partial_t \psi(x,t) = iv\partial_x \psi(x,t) + \frac{\Lambda'}{\sqrt{2\pi a}} f(t)\delta(x),$$

$$-i\partial_t \psi^\dagger(x,t) = iv\partial_x \psi^\dagger(x,t) - \frac{\Lambda'}{\sqrt{2\pi a}} f(t)\delta(x), \quad (28)$$

$$-i\partial_t f(t) = 2\frac{\Lambda'}{\sqrt{2\pi a}}[\psi(0,t) - \psi^\dagger(0,t)].$$

From this, one can map the problem into the scattering of the right-mover $\psi$ on a localized impurity. These equations are then solved by introducing a plane-wave decomposition of the fermionic field $\psi$ as

$$\psi(x,t) = \begin{cases} \int d\omega A_\omega e^{i\omega \frac{x}{v}} e^{-i\omega t} & \text{for } x < 0 \\ \int d\omega B_\omega e^{i\omega \frac{x}{v}} e^{-i\omega t} & \text{for } x > 0, \end{cases} \quad (29)$$

$$\psi^\dagger(x,t) = \begin{cases} \int d\omega A^\dagger_{-\omega} e^{i\omega \frac{x}{v}} e^{-i\omega t} & \text{for } x < 0 \\ \int d\omega B^\dagger_{-\omega} e^{i\omega \frac{x}{v}} e^{-i\omega t} & \text{for } x > 0. \end{cases} \quad (30)$$

Substituting these back into the equations of motion, using the definition $\psi(0) = [\psi(0^+) + \psi(0^-)]/2$ and integrating around the $\delta(x)$ function, one is left with

$$0 = iv \int d\omega (B_\omega - A_\omega) e^{-i\omega t} + \frac{\Lambda'}{\sqrt{2\pi a}} f(t),$$

$$0 = iv \int d\omega (B^\dagger_{-\omega} - A^\dagger_{-\omega}) e^{-i\omega t} - \frac{\Lambda'}{\sqrt{2\pi a}} f(t),$$

$$-i\partial_t f(t) = \frac{\Lambda'}{\sqrt{2\pi a}} \Bigg[ \int d\omega (B_\omega + A_\omega) e^{-i\omega t}$$

$$- \int d\omega (B^\dagger_{-\omega} + A^\dagger_{-\omega}) e^{-i\omega t} \Bigg]. \quad (31)$$

Combining these equations to get rid of $f$ and $B^\dagger$, one obtains

$$B_\omega = \frac{-i\omega}{\mathcal{T}_k - i\omega} A_\omega + \frac{\mathcal{T}_k}{\mathcal{T}_k - i\omega} A^\dagger_{-\omega}, \quad (32)$$

where we introduced the crossover energy scale

$$\mathcal{T}_k = \frac{4\pi a}{v}\left(\frac{\Lambda'}{2\pi a}\right)^2, \quad (33)$$

which is set by the tunneling amplitude $\Lambda'$. Following Ref. [57], the previous relation can be generalized for all filling factors taking the form

$$\mathcal{T}_k = \frac{2\omega_c}{g}\left(\frac{1}{2\Gamma(g)}\frac{\Lambda'}{a\omega_c}\right)^{\frac{1}{1-g}}. \quad (34)$$

Performing the Fourier transform back to time space one finally obtains the following relation between the Fourier components of the fermionic field $\psi$ before ($A$) and after ($B$) the QPC as

$$B(t) = A(t) - \mathcal{T}_k \int_{-\infty}^{t} e^{-\mathcal{T}_k(t-t')}[A(t') - A^\dagger(t')]. \quad (35)$$

In Ref. [36] and [34], all relevant transport quantities are then written down only in terms of averages of this newly defined fermionic $A$ field, which is free by construction. The corresponding propagator was naturally assumed to be trivially given by a Fermi function, corresponding to the reservoirs at equilibrium, i.e., equal temperature.

However, as we can see from Eq. (29), the new fermion $\psi$ is actually made of combined quasiparticles from the right and left reservoirs which, in our present case, correspond to different temperatures. One therefore needs to be particularly careful in expressing the propagator. One way of doing this is to revert to the bosonic description and to write $\psi$ in terms of the bosonic fields $\phi_R$ and $\phi_L$ taken at a position before the QPC and which are then uncoupled from each other. The full calculation of the propagator is detailed in Appendix A, here we report the final result for a (1/3, 1) junction which reads

$$\langle A^\dagger(t)A(t')\rangle = \langle \psi^\dagger(0^-,t)\psi(0^-,t')\rangle$$

$$= \frac{1}{2\pi a} e^{\frac{3}{4}\mathcal{G}_L(t-t')} e^{\frac{1}{4}\mathcal{G}_R(t-t')}, \quad (36)$$

where, since $A$ is only defined at a position before the QPC, the quantum averaging is performed on the state where the two edge states are decoupled and at their respective temperature $T_\alpha$. Here we also introduced the bosonic Green's function as [58,59]

$$\mathcal{G}_\alpha(\tau) = -\ln\left[\frac{\sinh\left(\pi T_\alpha\left(\frac{i}{\omega_c} - \tau\right)\right)}{\sinh\left(\frac{i}{\omega_c}\pi T_\alpha\right)}\right], \quad (37)$$

where $\omega_c = v/a$ is a high-energy cutoff and $T_\alpha$ is the temperature of the right or left QH bar.

In practice, we need two types of correlators, which we express from their Fourier transform as

$$\langle A^\dagger(t)A(t')\rangle = \int \frac{d\omega}{2\pi v} e^{i\omega(t-t')} n_\omega, \quad (38)$$

$$\langle A(t)A^\dagger(t')\rangle = \int \frac{d\omega}{2\pi v} e^{-i\omega(t-t')} (1 - n_\omega), \quad (39)$$

where according to Eq. (36), one has

$$n_\omega = \int d\tau e^{-i\omega\tau} \frac{\omega_c}{2\pi} e^{\frac{3}{4}\mathcal{G}_L(\tau)} e^{\frac{1}{4}\mathcal{G}_R(\tau)}. \quad (40)$$

Note that, although $n_\omega$ is not in all generality a Fermi distribution, the field $A$ is a fermionic field and thus satisfies the standard anticommutation relations.

### B. General expression of the delta-$T$ noise for (1/3, 1) junction

To evaluate the noise in Eq. (19) we need to consider the fluctuations of the tunneling current. The latter, in Eq. (8), is expressed in terms of the imbalance of the right densities before and after the QPC which are related to the Fourier components of the fermionic field $\psi$ before ($A$) and after ($B$) the QPC (see Appendix B for more details), namely,

$$I(t) = ev[\rho_R(0^+,t) - \rho_R(0^-,t)]$$

$$= \frac{ev}{2}[B^\dagger(t)B(t) + A^\dagger(t)A(t)]. \quad (41)$$

This expression can readily be understood from a current conservation perspective as the current tunneling between edges naturally corresponds to the difference of the current impinging on and the one outgoing from the QPC.

The fluctuations of the tunneling current are then readily given by

$$\mathcal{S}(t_1, t_2) = \langle \Delta I(t_1) \Delta I(t_2) \rangle$$
$$= \left(\frac{ev}{2}\right)^2 \langle [B^\dagger(t_1)B(t_1) + A^\dagger(t_1)A(t_1)]$$
$$\times [B^\dagger(t_2)B(t_2) + A^\dagger(t_2)A(t_2)]\rangle, \quad (42)$$

where one should only keep the connected contributions of the thermal average.

Since we are interested in the zero-frequency current noise given in Eq. (19), the final result reads

$$S(T_L, T_R) = 2 \int_{-\infty}^{+\infty} d\tau \mathcal{S}(0, \tau)$$
$$= 2\left(\mathcal{T}_k \frac{e}{2}\right)^2 \int \frac{d\omega}{2\pi} \Bigg\{ 2[n_\omega n_{-\omega}$$
$$+ (1 - n_\omega)(1 - n_{-\omega})] \left(\frac{\omega}{\mathcal{T}_k^2 + \omega^2}\right)^2$$
$$+ n_\omega(1 - n_\omega) \left(\frac{2}{\mathcal{T}_k} \frac{\omega^2}{\mathcal{T}_k^2 + \omega^2}\right)^2 \Bigg\}, \quad (43)$$

where all the detailed calculations are reported in Appendix C. It is worth noticing that, since we are considering the noise induced by a temperature difference, the distribution function $n_\omega$ depends on both temperatures as

$$n_\omega = \int d\tau e^{-i\omega\tau} \frac{\omega_c}{2\pi} \left[\frac{\sinh\left(\frac{i}{\omega_c}\pi T_L\right)}{\sinh\left(\pi T_L\left(\frac{i}{\omega_c} - \tau\right)\right)}\right]^{3/4}$$
$$\times \left[\frac{\sinh\left(\frac{i}{\omega_c}\pi T_R\right)}{\sinh\left(\pi T_R\left(\frac{i}{\omega_c} - \tau\right)\right)}\right]^{1/4}. \quad (44)$$

Using that $n_\omega + n_{-\omega} = 1$, the noise can be written under a much simpler form as

$$S(T_L, T_R) = e^2 \int \frac{d\omega}{2\pi} n_\omega(1 - n_\omega)\left(1 - \frac{\mathcal{T}_k^2 - \omega^2}{\mathcal{T}_k^2 + \omega^2}\right). \quad (45)$$

The noise can be obtained numerically for any set of temperatures $T_L, T_R$ and the information on the coupling strength is encoded in the energy crossover $\mathcal{T}_k$ which allows us to scan for the entire range of tunneling regimes.

In Fig. 3, we show the full delta-$T$ noise [as defined in Eq. (19)] as a function of the temperature difference, in unit of $\mathcal{T}_k$, for six different cuts. It comes out that in the weak- to moderate-temperature-bias regime, the leading contribution to the noise is linear in $\Delta T$, a feature that can be readily confirmed analytically, as we show in Appendix D. This leading linearly $\Delta T$ behavior survives even if one changes the temperature parametrization, introducing the average temperature $\bar{T} = (T_R + T_L)/2$. This may come as a surprise because it is in stark contrast with the results obtained for the homogeneous case ($\nu_R = \nu_L = \nu$) [23], where the dominant term was quadratic in the temperature difference. Indeed, in this latter situation, one is protected from such a linear contribution because of the symmetry of the system under the exchange of right and left leads, ensuring that only even terms in $\Delta T$ survive. The inhomogeneous case considered here breaks this symmetry, thus enabling linear contributions which then dominate the noise in the regime of weak temperature bias.

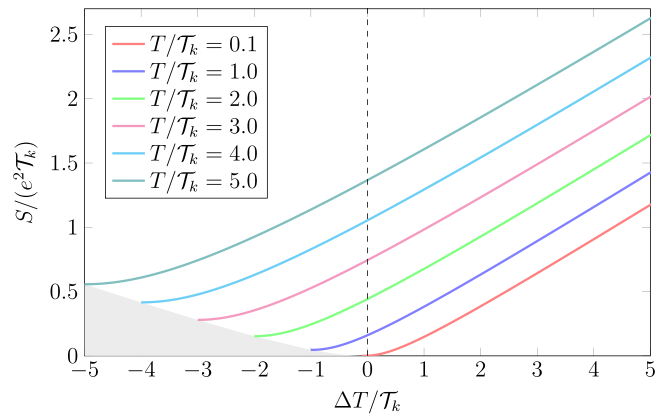

FIG. 3. Full noise $S$ for the (1/3, 1) junction, as a function of the temperature difference $\Delta T$ (given in units of the energy scale $\mathcal{T}_k$) for different values of the right lead temperature, $T/\mathcal{T}_k = 0.1, 1, 2, 3, 4$ and 5. The gray area corresponds to the unphysical region where $T_L = T + \Delta T < 0$.

Notice that, in presence of an extended contact, disorder effects could lead to asymmetries also in the homogeneous case. This could lead to a noise contribution linear in the temperature mismatch. However, we expect this contribution to be sample dependent and not universally related to the scaling dimension of the tunneling operators differently from the one we are investigating (see below). From the experimental standpoint, the emergence of a linear correction as a function of the temperature difference constitutes a major improvement compared with the homogeneous case because it makes the delta-$T$ noise a lot easier to probe.

In this section we have been able to solve exactly the problem of tunneling between a QH junction with the only constraint on filling factors, being $\nu_L = 1/3$ and $\nu_R = 1$. This interesting result can be enriched by extending it to a general dependence of the noise on filling factors without specifying their values *a priori*. Moreover, in addition to the numerical evaluation, it could be interesting to work out an analytic expression for the delta-$T$ noise. As we see below, this can be achieved for comparable temperatures of the two QH bars, as this approach relies on an expansion in the temperature difference $\Delta T$. Nevertheless, considering small temperature differences allows us to be closer to the experimentally realistic situation, where large temperature gradients between reservoirs are difficult to control.

## VI. UNIVERSAL EXPRESSION FOR DELTA-$T$ NOISE

In this section, we derive a universal formula for the first-order expansion of the noise in Eq. (19) that applies to all orders in the tunneling amplitude $\Lambda$ and for any set of filling factors $(\nu_L, \nu_R)$.

Since we are considering different temperatures between the two QH bars, parametrized as in Eq. (20), and according to the rotations introduced in previous sections, we can expand the noise $S(T_L, T_R)$ up to the first order in the ratio $\Delta T/T$.

Following the calculations given in Appendix E, one has

$$S(T_L, T_R) = S_0(T) + \frac{\Delta T}{T^2} \int_{-\infty}^{+\infty} d\tau \frac{1}{Z} \text{Tr}\{e^{-\beta(H_+^{(0)}+H_-^{(0)})}$$
$$\times H_{+,-} I(\varphi_-(\tau))I(\varphi_-(0))\} + O(\Delta T^2), \quad (46)$$

where $\beta = T^{-1}$ and the mixed term $H_{+,-}$ due to the expansion in $\Delta T$ is given by

$$H_{+,-} = \frac{v}{4\pi} \int dx \left[ \frac{\cos^2\theta}{2} (\partial_x\varphi_+ + \partial_x\varphi_-)^2 \right.$$
$$- \sin\theta\cos\theta(\partial_x\varphi_+ + \partial_x\varphi_-)(\partial_x\varphi_+ - \partial_x\varphi_-)$$
$$\left. + \frac{\sin^2\theta}{2}(\partial_x\varphi_+ - \partial_x\varphi_-)^2 \right]. \quad (47)$$

In Eq. (46), $S_0(T)$ is the equilibrium noise at $\Delta T = 0$, namely,

$$S_0(T) = S(T_L, T_R)\Big|_{\Delta T=0}$$
$$= \frac{1}{Z_+^{(0)} Z_-^{(0)}} \int_{-\infty}^{+\infty} d\tau$$
$$\times \text{Tr}\{e^{-\beta(H_+^{(0)}+H_-^{(0)})} I(\varphi_-(\tau))I(\varphi_-(0))\}, \quad (48)$$

with

$$Z_\pm^{(0)} = \text{Tr}\{e^{-\beta H_\pm^{(0)}}\}. \quad (49)$$

Furthermore, in Eq. (46) one needs to consider also the first-order expansion of $Z$ which reads

$$Z = \text{Tr}\left\{ e^{-\beta(H_+^{(0)}+H_-^{(0)})} \left(1 + \frac{\Delta T}{T^2} H_{+,-}\right) \right\}. \quad (50)$$

Finally, putting everything together, the delta-$T$ noise of Eq. (19), expanded to first order in $\Delta T$, reads

$$S(T_L, T_R) = S_0(T) + \Sigma(\nu_L, \nu_R, T)\Delta T + O(\Delta T^2), \quad (51)$$

with

$$\Sigma(\nu_L, \nu_R, T) = -\left(\frac{\nu_R}{\nu_R + \nu_L}\right) \frac{1}{T^2} \frac{\partial S_0}{\partial \beta}. \quad (52)$$

We underline the relevance of this result which enables to calculate the first-order correction to the noise in the temperature gradient only by knowing the expression for the equilibrium noise $S_0(T)$. In particular, our derivation does not require any assumption concerning the strength of the tunneling between the two QH bars. This allows us to obtain the out-of-equilibrium delta-$T$ noise in various tunneling regimes, provided that one is able to compute the corresponding thermal noise at equilibrium. Since Eqs. (51) and (52) are valid for all values of $\Lambda$, it is worth noticing that they can be exploited for describing both the weak-coupling regime and the dual strong-coupling model.

## VII. EXPLICIT RESULTS FOR DELTA-$T$ NOISE

In the previous section, we show that it is possible to obtain a universal expression for the delta-$T$ noise at first order in $\Delta T/T$ without specifying the tunneling strength between the two QH systems.

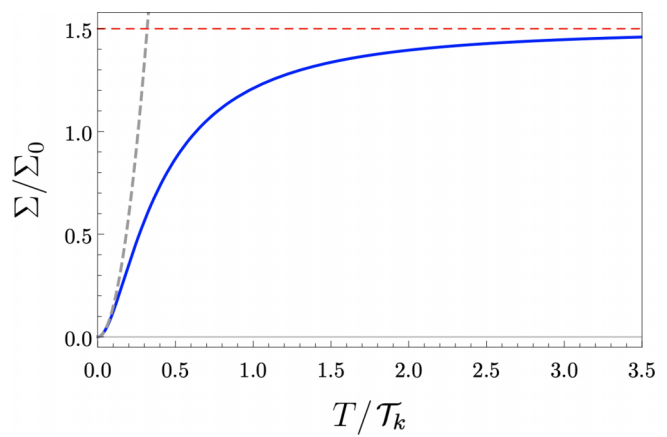

FIG. 4. Behavior of the linear coefficient $\Sigma \equiv \Sigma(1/3, 1, T)$ in Eq. (54) in units of $\Sigma_0 = e^2/2\pi$ as a function of $T/\mathcal{T}_k$. The dashed gray line stands for the weak-coupling limit [see Eq. (55)] when $T \ll \mathcal{T}_k$ ($\Lambda \to 0$, $\Lambda' \to \infty$). The horizontal dashed red line describes the strong-coupling limit for $T \gg \mathcal{T}_k$ ($\Lambda \to \infty$, $\Lambda' \to 0$) [see Eq. (56)].

Here, we investigate in detail the two opposite regimes of weak and strong coupling between the two Hall fluids. We start by applying our universal formula, Eq. (51), to the specific case of the ($\nu_L = 1/3$, $\nu_R = 1$) junction. There, an exact expression of the equilibrium noise can be derived at all orders in the tunnel coupling [36], following the refermionization procedure introduced in the previous section. Then, we generalize this approach for a system with generic filling factor combinations, concentrating on the two opposite regimes of the coupling strength. It is worth emphasizing that, although our focus is on the weak- and strong-coupling regimes, in principle, our approach allows us to calculate the full noise for any value of the tunneling parameter, provided that we have the corresponding expression of the equilibrium noise.

### A. Exact solution for a $(\frac{1}{3}, 1)$ junction

By exploiting the exact refermionization procedure, given in the previous section, we are able to evaluate the equilibrium noise $S_0(T)$ in the case ($\nu_L = 1/3$, $\nu_R = 1$), recovering the result from Ref. [36] at zero voltage. The final expression, whose derivation is given in Appendix F, reads

$$S_0(T) = \frac{1}{2}\left(\frac{e^2}{2\pi}\right)\mathcal{T}_k\left[4\frac{T}{\mathcal{T}_k} - \frac{2}{\pi}\psi'\left(\frac{1}{2} + \frac{\mathcal{T}_k}{2\pi T}\right)\right], \quad (53)$$

with $\mathcal{T}_k$ defined in Eq. (34). Then, according to Eq. (52), one can write the coefficient of the first-order correction to the delta-$T$ noise:

$$\Sigma\left(\frac{1}{3}, 1, T\right) = \left(\frac{e^2}{2\pi}\right)\left[\frac{3}{2} + \frac{3}{8\pi^2}\psi''\left(\frac{1}{2} + \frac{\mathcal{T}_k}{2\pi T}\right)\left(\frac{\mathcal{T}_k}{T}\right)^2\right], \quad (54)$$

with $\psi''$ being the second derivative of the digamma function. The behavior of this quantity as a function of $T/\mathcal{T}_k$ is reported in Fig. 4.

We can now consider two interesting limits of Eq. (54): the weak-coupling limit described by $\Lambda \to 0$ or equivalently, due to Eq. (23), $\Lambda' \to \infty$ [i.e., $\mathcal{T}_k \to \infty$ from Eq. (33)] and the strong-coupling limit where $\Lambda \to \infty$ and consequently $\Lambda'$ and $\mathcal{T}_k$ go to zero. This means that, in terms of the ratio $T/\mathcal{T}_k$, in the weak-coupling regime ($T \ll \mathcal{T}_k$) one has

$$\Sigma^{(wc)}\left(\frac{1}{3}, 1, T\right) = \left(\frac{e^2}{2\pi}\right)\frac{3\pi^2}{2}\left(\frac{T}{\mathcal{T}_k}\right)^2, \quad (55)$$

while in the strong-coupling limit ($T \gg \mathcal{T}_k$) one obtains

$$\Sigma^{(sc)}\left(\frac{1}{3}, 1, T\right) = \left(\frac{e^2}{2\pi}\right)\frac{3}{2}. \quad (56)$$

It is worth noting that Eqs. (55) and (56) are related to the slope of the curves shown in Fig. 3 near the value of $\Delta T = 0$. When the ratio $T/\mathcal{T}_k$ is small, meaning that we are considering the weak-coupling limit, we see that the plot has a nonlinear behavior, near $\Delta T = 0$, reflecting the quadratic dependence of $\Sigma^{(wc)}$ on $T/\mathcal{T}_k$ in Eq. (55). However, for the strong-coupling regime, where bigger ratios $T/\mathcal{T}_k$ have been considered, the slope of the plot is linear according to Eq. (56), where $\Sigma^{(sc)}$ is independent of $T/\mathcal{T}_k$.

Moreover, in Fig. 4 we observe the behavior of the linear coefficient in Eq. (54) for a $(1/3, 1)$ junction in terms of the ratio between temperature $T$ and the energy scale $\mathcal{T}_k$, related to the tunneling amplitude. This coefficient vanishes at zero temperature then evolves quadratically, consistently with Eq. (55). As temperature increases further, $\Sigma$ continuously increases before saturating as it gets closer to the limiting value of $3/2$ (in units of $e^2/2\pi$) which corresponds to the strong-coupling value of Eq. (56). This latter value represents the maximum reachable one for the linear order correction in the considered case.

### B. Perturbative regimes

In this section, we write the explicit expression for the equilibrium noise $S_0(T)$ due to the tunneling current, starting from the weak-coupling regime for all filling factors $\nu_L$ and $\nu_R$. We then switch to the strong-coupling limit by exploiting the duality relation discussed in Sec. IV.

In the weak-coupling regime, the explicit form for the equilibrium noise $S_0(T)$ is obtained through a perturbative expansion up to the second order in the tunneling Hamiltonian, Eq. (11), by using the Keldysh formalism [58,60]

$$S_0^{(wc)}(T) = \left(\frac{e\Lambda}{\pi a}\right)^2 \int_{-\infty}^{+\infty} d\tau e^{\frac{2}{g}\mathcal{G}(\tau)}, \quad (57)$$

with the bosonic Green's function given in Eq. (37) with $T_\alpha = T$ and $g = 2(\nu_L \nu_R)/(\nu_L + \nu_R)$.

Then the equilibrium noise in the weak-coupling regime reads [59,61,62]

$$S_0^{(wc)}(T) = \left(\frac{e^2}{2\pi}\right)\left(\frac{2\Lambda}{\omega_c a}\right)^2 \left(\frac{2\pi}{\omega_c}\right)^{\frac{2}{g}-2} T^{\frac{2}{g}-1} \frac{\Gamma^2\left(\frac{1}{g}\right)}{\Gamma\left(\frac{2}{g}\right)}. \quad (58)$$

In the case of free fermions $\nu_L = \nu_R = 1$, Eq. (58) yields the Johnson-Nyquist linearity of the equilibrium noise as a function of temperature. Furthermore, if we consider equal filling factors $\nu_L = \nu_R$ we recover the result reported in Ref. [23] for the thermal noise calculated in the tunneling regime. It is worth noticing that expression (58) only depends on $g$ which is the sum of the inverse filling factors [36] which does not allow us to discriminate between configurations such as ($\nu_R = 1$, $\nu_L = 1/5$) or ($\nu_R = \nu_L = 1/3$) only by looking at the equilibrium noise.

The equilibrium noise of Eq. (58) in the weak-coupling regime for all filling factors can be rewritten in terms of the crossover energy scale instead of the tunneling amplitude by exploiting Eqs. (23) and (34), namely

$$S_0^{(wc)}(T) = \frac{1}{4}\left(\frac{e^2}{2\pi}\right)(4\pi)^{\frac{2}{g}-2}\frac{\Gamma^4\left(\frac{1}{g}\right)}{\Gamma\left(\frac{2}{g}\right)}\left(\frac{T}{\mathcal{T}_k}\right)^{\frac{2}{g}-1}\mathcal{T}_k, \quad (59)$$

and consequently from Eq. (52) the linear coefficient for a generic $(\nu_L, \nu_R)$ junction is

$$\Sigma^{(wc)}(\nu_L, \nu_R, T) = \frac{1}{4}\left(\frac{e^2}{2\pi}\right)\left(\frac{\nu_R}{\nu_R + \nu_L}\right)\left(\frac{2}{g} - 1\right) \\ \times (4\pi)^{\frac{2}{g}-2}\frac{\Gamma^4\left(\frac{1}{g}\right)}{\Gamma\left(\frac{2}{g}\right)}\left(\frac{T}{\mathcal{T}_k}\right)^{\frac{2}{g}-2}. \quad (60)$$

Notice that for $\nu_R = 1$ and $\nu_L = 1/3$, this last equation reduces to the result derived in Eq. (55) for the weak-coupling limit $T \ll \mathcal{T}_k$.

Now, the strong-coupling limit can be studied using the weak-strong duality transformation, introduced in Sec. IV, where we consider the electron-quasiparticle substitution for the charge $e \to e^* = ge$ and for the filling factor $g \to 1/g$. By focusing on the strong-coupling limit we consider $\Lambda'$ instead of $\Lambda$ as depicted in Fig. 2(d). According to this, Eq. (58) leads to

$$S_B^{(sc)}(T) = \left(\frac{e^{*2}}{2\pi}\right)\left(\frac{2\Lambda'}{\omega_c a}\right)^2 \left(\frac{2\pi}{\omega_c}\right)^{2g-2} T^{2g-1} \frac{\Gamma^2(g)}{\Gamma(2g)}. \quad (61)$$

It is worth mentioning that the weak-strong duality transformation amounts to consider a dual system which can be treated in the weak-coupling limit, i.e., perturbatively in $\Lambda'$. According to this, at equilibrium, the tunneling current noise maps into a backscattered current noise, hence the notation $S_B$.

Using the expression for $\mathcal{T}_k$ from Eq. (34), we can rewrite the equilibrium backscattering noise in the strong-coupling regime for all filling factors as

$$S_B^{(sc)}(T) = (4g)^2 \left(\frac{e^2}{2\pi}\right)\left(\frac{4\pi}{g}\right)^{2g-2}\frac{\Gamma^4(g)}{\Gamma(2g)}\left(\frac{T}{\mathcal{T}_k}\right)^{2g-1}\mathcal{T}_k. \quad (62)$$

The noise associated with the tunneling current is then readily obtained from its backscattering counterpart by accounting for a bare equilibrium noise contribution linear in temperature which ends up dominating the transport for temperatures $T \gg \mathcal{T}_k$.

In the end, it leads to the total equilibrium tunneling noise in the strong-coupling regime

$$S_0^{(sc)}(T) = \left(\frac{e^2}{2\pi}\right)4gT - S_B^{(sc)}(T). \quad (63)$$

In this case, the linear coefficient of the $\Delta T$ noise can be written as

$$\Sigma^{(sc)}(\nu_L, \nu_R, T) = \left(\frac{e^2}{2\pi}\right)\left(\frac{\nu_R}{\nu_R + \nu_L}\right)\left[4g + (4g)^2(1-2g)\right.$$
$$\left. \times \left(\frac{4\pi}{g}\right)^{2g-2}\frac{\Gamma^4(g)}{\Gamma(2g)}\left(\frac{\mathcal{T}_k}{T}\right)^{2-2g}\right], \quad (64)$$

which then reduces to Eq. (56) for the (1, 1/3) junction.

Notice that the analysis carried out so far is proper of the case of inhomogeneous Hall junction with filling factors belonging to the Laughlin sequence. Here, clearly emerge the role played by the scaling dimension associated with the different FQH states involved in the tunneling (related to the filling factors $\nu_L$ and $\nu_R$). In more general composite FQH states the emergence of neutral modes affects the scaling dimension of the tunneling operators and need to be carefully taken into account [61].

As a final remark we notice that the first-order contribution to the delta-$T$ noise in the temperature bias cannot be solely expressed in terms of the effective filling factor $g$. This is true for both the weak- and strong-coupling regimes as can be readily seen from Eqs. (60) and (64). Since the linear in $\Delta T$ coefficient depends separately on $\nu_L$ and $\nu_R$, this specific signature allows us to distinguish between different filling factor combinations which nevertheless have the same effective filling factor $g$ and consequently the same equilibrium noise, such as ($\nu_R = 1$, $\nu_L = 1/5$) or ($\nu_R = \nu_L = 1/3$). This further highlights the importance of the low-order delta-$T$ noise as a relevant probe of the transport mechanisms at play in the general ($\nu_R$, $\nu_L$) junction.

## VIII. CONCLUSION

This work was devoted to the study of the nonequilibrium noise generated by a temperature gradient between two FQH systems, known as delta-$T$ noise.

We have considered an inhomogeneous QH junction and we have demonstrated the predominant contribution to the noise is linear in the temperature gradient, differently from a homogeneous junction where the first nonzero contribution is quadratic [23].

Moreover, we have considered the two Hall bars characterized by strong interaction, focusing on a coupling between the two edges whose intensity can be either considered in a weak or strong-coupling regime thanks to a weak-strong duality transformation.

We have solved exactly the problem of the delta-$T$ noise for the (1/3, 1) junction, demonstrating that this regime enables to explore the full range of tunnel coupling and to consider any set of temperatures $T_R$ and $T_L$ without restrictions.

In addition, we have reported on a universal expression, in terms of the tunneling parameter for a completely generic junction, for the linear correction to the full delta-$T$ noise in the temperature gradient starting from the knowledge of the equilibrium noise.

Furthermore, we have demonstrated that we are able again to cross from the weak-coupling regime to the strong-coupling regime by applying a duality transformation and we have reported on the asymptotic behavior of the linear coefficient of this expansion for generic values of the filling factors ($\nu_L$, $\nu_R$).

Finally, our approach shows the relevance of delta-$T$ noise in better understanding the transport properties of such strongly correlated systems, unlike previously considered noise contributions, since it depends on both filling factors separately rather than the sole effective filling factor describing the junction.

This work offers many interesting perspectives, essentially related to practical realizations of such temperature-biased inhomogeneous junctions. Indeed, junctions between different Hall fluids are notoriously difficult to implement experimentally and the careful investigation of their transport properties remain largely unexplored. As the magnetic field is constant everywhere across the two-dimensional electron gas, regions of different filling factors require different electron density, which is typically achieved via electrostatic gates whose close proximity leads to a severe risk of shorting each other. Several solutions have been envisioned to circumvent this issue. However, these raise several challenging problems for theoretical modeling.

One aspect that could be explored in forthcoming studies is the effect of local charge depletion at the QPC, as a consequence of electrostatic effects. This leads to a local filling factor in the region of the point contact, whose importance for transport properties has been previously underlined [63,64].

Moreover, interesting new perspectives in this field could be opened by the study of composite FQH edge state, where the emergence of co- and counterpropagating neutral modes could complicate the presented picture [44,61,65,66] or more exotic states such as fractional Chern insulators [67,68].

Another fascinating direction to explore is the case of long junctions, where Andreev reflection-like processes have been observed recently [32]. This would theoretically involve multiple, randomly distributed, quantum point contacts and brings about the importance of coherence effects across these.

## ACKNOWLEDGMENTS

Authors G.R., D.F. and M.S. acknowledge the support of the "Dipartimento di Eccellenza MIUR 2018-22" (Università di Genova, DIFI). D.F. also wants to acknowledge the financial support of the Second Scientific Label 2021 of the "Université Franco-Italienne." This work received support from the French government under the France 2030 investment plan, as part of the Initiative d'Excellence d'Aix-Marseille Université - A*MIDEX (authors J.R., T.J. and T.M.).

## APPENDIX A: EXPLICIT EVALUATION OF THE PROPAGATOR

In this Appendix, we compute explicitly the propagator $\langle A^\dagger(t)A(t')\rangle$ of Eq. (36). The latter only involves the fermionic field $A$, which is only defined at a position before the QPC, so that the quantum averaging is performed on the state where the two edge states are decoupled and at their respective temperature $T_R$ and $T_L$.

We first recall the relation in Eq. (21) between the rotated fields $\varphi_R$ and $\varphi_L$ and the dual ones $\widetilde{\varphi}_R$ and $\widetilde{\varphi}_L$. The idea in order to evaluate the propagator is to revert to the bosonic description, where

$$
\begin{aligned}
\langle A^{\dagger}(t) A(t') \rangle &= \langle \psi^{\dagger}(0^-, t) \psi(0^-, t') \rangle \\
&= \frac{1}{2\pi a} \langle e^{-i\widetilde{\varphi}_-(0^-,t)} e^{i\widetilde{\varphi}_-(0^-,t')} \rangle \\
&= \frac{1}{2\pi a} \langle e^{-\frac{i}{\sqrt{2}}\widetilde{\varphi}_L(0^-,t) - \widetilde{\varphi}_R(0^-,t)} e^{\frac{i}{\sqrt{2}}\widetilde{\varphi}_L(0^-,t') - \widetilde{\varphi}_R(0^-,t')} \rangle \\
&= \frac{1}{2\pi a} \langle e^{-\frac{i}{\sqrt{2}}\varphi_L(0^-,t) - \varphi_R(0^-,t)} e^{\frac{i}{\sqrt{2}}\varphi_L(0^-,t') - \varphi_R(0^-,t')} \rangle \\
&= \frac{1}{2\pi a} \langle e^{-i[\frac{\cos\theta + \sin\theta}{\sqrt{2}} \phi_L(0^-,t) + \frac{\sin\theta - \cos\theta}{\sqrt{2}} \phi_R(0^-,t)]} \\
&\quad \times e^{i[\frac{\cos\theta + \sin\theta}{\sqrt{2}} \phi_L(0^-,t') + \frac{\sin\theta - \cos\theta}{\sqrt{2}} \phi_R(0^-,t')]} \rangle. \quad (A1)
\end{aligned}
$$

The bosonic fields $\phi_R$ and $\phi_L$ are taken at a position before the QPC ($x = 0^-$) and are thus uncoupled from each other. For this reason we are able to rewrite the correlator as a product of averages evaluated at a fixed temperature $T_R$ and $T_L$ of the considered right or left QH bar. Then, for the particular case of a QH junction with $\nu_L = 1/3$ and $\nu_R = 1$ we recover the result of Eq. (36) in the main text, namely,

$$
\begin{aligned}
\langle A^{\dagger}(t) A(t') \rangle &= \frac{1}{2\pi a} \langle e^{-i\frac{\sqrt{3}}{2}\phi_L(0^-,t)} e^{i\frac{\sqrt{3}}{2}\phi_L(0^-,t')} \rangle \\
&\quad \times \langle e^{i\frac{1}{2}\phi_R(0^-,t)} e^{-i\frac{1}{2}\phi_R(0^-,t')} \rangle \\
&= \frac{1}{2\pi a} e^{\frac{3}{4}\mathcal{G}_L(t-t')} e^{\frac{1}{4}\mathcal{G}_R(t-t')}. \quad (A2)
\end{aligned}
$$

## APPENDIX B: PARTICLE DENSITY RELATIONS

In this Appendix, we give a description of the junction in terms of the particle densities of the edge channels and we restore the equivalence in Eq. (41). This allows us to compute the current noise for the specific case of (1/3, 1) presented in Sec. V B.

The particle densities $\rho_R$ and $\rho_L$ of the two edges of the inhomogeneous QH junction are defined in Eq. (2) and depend on the fields $\phi_L$ and $\phi_R$. The first change of basis, which relates $\phi_L$ and $\phi_R$ to $\varphi_L$ and $\varphi_R$ [see Eq. (9)], can be expressed in terms of the particle densities as follows:

$$
\begin{pmatrix} \rho_L \\ \rho_R \end{pmatrix} = \begin{pmatrix} \frac{\sqrt{\nu_L} + \sqrt{\nu_R}}{2} & \frac{\sqrt{\nu_L} - \sqrt{\nu_R}}{2} \\ \frac{\sqrt{\nu_R} - \sqrt{\nu_L}}{2\sqrt{\nu_L \nu_R}} & \frac{\sqrt{\nu_R} + \sqrt{\nu_L}}{2\sqrt{\nu_L \nu_R}} \end{pmatrix} \begin{pmatrix} \rho'_L \\ \rho'_R \end{pmatrix}. \quad (B1)
$$

After the introduction of the second rotation of Eq. (13), the new densities $\rho_+$ and $\rho_-$ are related to $\rho'_L$ and $\rho'_R$ by

$$
\begin{pmatrix} \rho'_L \\ \rho'_R \end{pmatrix} = \sqrt{\frac{g}{2}} \begin{pmatrix} 1 & 1 \\ 1 & -1 \end{pmatrix} \begin{pmatrix} \rho_+ \\ \rho_- \end{pmatrix}. \quad (B2)
$$

In terms of densities, the duality relates $\rho'_L$ and $\rho'_R$ to the dual ones $\widetilde{\rho}_L$ and $\widetilde{\rho}_R$. Notice, from the dual fields transformation in Eq. (21), that the densities in the incoming channels $\rho'_L(x < 0)$ and $\rho'_R(x < 0)$ are the same as $\widetilde{\rho}_L(x < 0)$ and $\widetilde{\rho}_R(x < 0)$ [i.e., from Eq. (21) for $x < 0$ we have $\varphi_L = \widetilde{\varphi}_L$ and $\varphi_R = \widetilde{\varphi}_R$ and consequently for the densities]. Furthermore, the matrices given in Eqs. (B1) and (B2) can be used to express the original fields $\phi_L(x < 0)$ and $\phi_R(x < 0)$ in terms of the fields $\widetilde{\varphi}_+(x < 0)$ and $\widetilde{\varphi}_-(x < 0)$.

To write the original densities in the outgoing channels $\rho_L(x > 0)$ and $\rho_R(x > 0)$ in terms of the densities $\widetilde{\rho}_+(x > 0)$ and $\widetilde{\rho}_-(x > 0)$ it is necessary to realize that the duality transformation of Eq. (21) exchanges $\varphi_L$ and $\varphi_R$ for $x > 0$. As a consequence, for $x > 0$ Eq. (B1) reads

$$
\begin{pmatrix} \rho_L \\ \rho_R \end{pmatrix} = \begin{pmatrix} \frac{\sqrt{\nu_L} + \sqrt{\nu_R}}{2} & \frac{\sqrt{\nu_L} - \sqrt{\nu_R}}{2} \\ \frac{\sqrt{\nu_R} - \sqrt{\nu_L}}{2\sqrt{\nu_L \nu_R}} & \frac{\sqrt{\nu_R} + \sqrt{\nu_L}}{2\sqrt{\nu_L \nu_R}} \end{pmatrix} \begin{pmatrix} \widetilde{\rho}_R \\ \widetilde{\rho}_L \end{pmatrix}. \quad (B3)
$$

It is thus useful to connect the corresponding $\pm$ density operators with their initial $(L, R)$ counterparts. By using Eqs. (B1)–(B3) the densities $\rho_R$ and $\rho_L$ can be written in terms of $\widetilde{\rho}_+$ and $\widetilde{\rho}_-$,

$$
\rho_L(x = 0^-, t) = \sqrt{\frac{g\nu_L}{2}} \widetilde{\rho}_+(0^-, t) + \sqrt{\frac{g\nu_R}{2}} \widetilde{\rho}_-(0^-, t),
$$

$$
\rho_R(x = 0^-, t) = \sqrt{\frac{g}{2\nu_L}} \widetilde{\rho}_+(0^-, t) - \sqrt{\frac{g}{2\nu_R}} \widetilde{\rho}_-(0^-, t),
$$
(B4)

and

$$
\rho_L(x = 0^+, t) = \sqrt{\frac{g\nu_L}{2}} \widetilde{\rho}_+(0^+, t) - \sqrt{\frac{g\nu_R}{2}} \widetilde{\rho}_-(0^+, t),
$$

$$
\rho_R(x = 0^+, t) = \sqrt{\frac{g}{2\nu_L}} \widetilde{\rho}_+(0^+, t) + \sqrt{\frac{g}{2\nu_R}} \widetilde{\rho}_-(0^+, t).
$$
(B5)

Since the $\widetilde{\varphi}_+$ field is continuous through the QPC, we can drop the $\widetilde{\rho}_+$ contribution and the current operator is therefore given by

$$
I(t) = ev \sqrt{\frac{g}{2}} [\widetilde{\rho}_-(0^+, t) + \widetilde{\rho}_-(0^-, t)]. \quad (B6)
$$

Exploiting the decomposition in Eqs. (29) and (30) for the specific case of the (1/3, 1) junction (i.e., $g = 1/2$), the tunneling current can be rewritten in terms of the Fourier components of the fermionic field before ($A$) and after ($B$) the QPC, leading to

$$
I(t) = \frac{ev}{2} [A^{\dagger}(t) A(t) + B^{\dagger}(t) B(t)]. \quad (B7)
$$

This last equation is thus an equivalent definition of the current operator introduced in Eq. (5) and it is quoted in the main text as Eq. (41).

## APPENDIX C: ZERO-FREQUENCY NOISE FOR A MISMATCHED (1/3, 1) QUANTUM HALL JUNCTION

In this Appendix we derive the result of Eq. (43). To do this, we recall Eq. (35) where the fermionic fields $B$ are written in terms of the $A$ ones so that one has

$$B^\dagger(t)B(t) + A^\dagger(t)A(t) = -\mathcal{T}_k \int_{-\infty}^t dt' e^{-\mathcal{T}_k(t-t')} A^\dagger(t)[A(t') - A^\dagger(t')] - \mathcal{T}_k \int_{-\infty}^t dt' e^{-\mathcal{T}_k(t-t')}[A^\dagger(t') - A(t')]A(t) + 2A^\dagger(t)A(t)$$

$$+ \mathcal{T}_k^2 \int_{-\infty}^t dt_1' \int_{-\infty}^t dt_2' e^{-\mathcal{T}_k(t-t_1')} e^{-\mathcal{T}_k(t-t_2')} \{A^\dagger(t_1'), A(t_2')\}$$

$$= \frac{\mathcal{T}_k}{2v} + \mathcal{T}_k \int_{-\infty}^t dt' e^{-\mathcal{T}_k(t-t')}[A^\dagger(t)A^\dagger(t') - A^\dagger(t)A(t') + A(t')A(t) - A^\dagger(t')A(t) + 2A^\dagger(t)A(t)]. \quad (C1)$$

Substituting this expression back into the definition for the noise, Eq. (42), one has

$$\mathcal{S}(t_1, t_2) = \left(\mathcal{T}_k \frac{ev}{2}\right)^2 \int_{-\infty}^{t_1} dt_1' \int_{-\infty}^{t_2} dt_2' e^{-\mathcal{T}_k(t_1 - t_1' + t_2 - t_2')} \langle [A^\dagger(t_1)A^\dagger(t_1') - A^\dagger(t_1)A(t_1') + A(t_1')A(t_1)$$

$$- A^\dagger(t_1')A(t_1) + 2A^\dagger(t_1)A(t_1)][A^\dagger(t_2)A^\dagger(t_2') - A^\dagger(t_2)A(t_2') + A(t_2')A(t_2) - A^\dagger(t_2')A(t_2) + 2A^\dagger(t_2)A(t_2)]\rangle, \quad (C2)$$

which becomes, after applying Wick's theorem,

$$\mathcal{S}(t_1, t_2)$$
$$= \left(\mathcal{T}_k \frac{ev}{2}\right)^2 \int_{-\infty}^{t_1} dt_1' \int_{-\infty}^{t_2} dt_2' e^{-\mathcal{T}_k(t_1 - t_1' + t_2 - t_2')} \{\langle A^\dagger(t_1)A(t_2)\rangle\langle A^\dagger(t_1')A(t_2')\rangle - \langle A^\dagger(t_1)A(t_2')\rangle\langle A^\dagger(t_1')A(t_2)\rangle$$
$$+ \langle A^\dagger(t_1)A(t_2')\rangle\langle A(t_1')A^\dagger(t_2)\rangle + \langle A^\dagger(t_1)A(t_2)\rangle\langle A(t_1')A^\dagger(t_2')\rangle + \langle A(t_1')A^\dagger(t_2')\rangle\langle A(t_1)A^\dagger(t_2)\rangle - \langle A(t_1')A^\dagger(t_2)\rangle\langle A(t_1)A^\dagger(t_2')\rangle$$
$$+ \langle A^\dagger(t_1')A(t_2')\rangle\langle A(t_1)A^\dagger(t_2)\rangle + \langle A^\dagger(t_1')A(t_2)\rangle\langle A(t_1)A^\dagger(t_2')\rangle - 2[\langle A^\dagger(t_1)A(t_2)\rangle\langle A(t_1')A^\dagger(t_2)\rangle + \langle A^\dagger(t_1')A(t_2)\rangle\langle A(t_1)A^\dagger(t_2)\rangle$$
$$+ \langle A^\dagger(t_1)A(t_2')\rangle\langle A(t_1)A^\dagger(t_2)\rangle + \langle A^\dagger(t_1)A(t_2)\rangle\langle A(t_1)A^\dagger(t_2')\rangle] + 4\langle A^\dagger(t_1)A(t_2)\rangle\langle A(t_1)A^\dagger(t_2)\rangle\}. \quad (C3)$$

Using the Fourier-transformed versions of the correlators from Eqs. (38) and (39), this is further rewritten as

$$\mathcal{S}(t_1, t_2) = \left(\mathcal{T}_k \frac{ev}{2}\right)^2 \int \frac{d\omega_1}{2\pi v} \int \frac{d\omega_2}{2\pi v} \left\{ e^{i(\omega_1 + \omega_2)(t_1 - t_2)} n_{\omega_1} n_{\omega_2} \left[\frac{1}{\mathcal{T}_k - i\omega_2} - \frac{1}{\mathcal{T}_k - i\omega_1}\right] \frac{1}{\mathcal{T}_k + i\omega_2}\right.$$

$$+ e^{-i(\omega_1 + \omega_2)(t_1 - t_2)}(1 - n_{\omega_1})(1 - n_{\omega_2}) \left[\frac{1}{\mathcal{T}_k + i\omega_1} - \frac{1}{\mathcal{T}_k + i\omega_2}\right]\frac{1}{\mathcal{T}_k - i\omega_1}$$

$$+ e^{i(\omega_1 - \omega_2)(t_1 - t_2)} n_{\omega_1}(1 - n_{\omega_2}) \left[\frac{1}{\mathcal{T}_k - i\omega_1} + \frac{1}{\mathcal{T}_k + i\omega_2}\right] \frac{1}{\mathcal{T}_k - i\omega_2} + e^{i(\omega_1 - \omega_2)(t_1 - t_2)} n_{\omega_1}(1 - n_{\omega_2})$$

$$\left. \times \left[\frac{1}{\mathcal{T}_k - i\omega_1} + \frac{1}{\mathcal{T}_k + i\omega_2}\right]\frac{1}{\mathcal{T}_k + i\omega_1} - 2e^{i(\omega_1 - \omega_2)(t_1 - t_2)} n_{\omega_1}(1 - n_{\omega_2})\left[\frac{2}{\mathcal{T}_k^2 + \omega_2^2} + \frac{2}{\mathcal{T}_k^2 + \omega_1^2} - \frac{2}{\mathcal{T}_k^2}\right]\right\}, \quad (C4)$$

where we also performed the time integrals.

Since we want to consider the noise power $S(T_L, T_R)$ we need to perform an additional time integration over $t_1 - t_2$, leading to

$$S(T_L, T_R) = 2\left(\mathcal{T}_k \frac{e}{2}\right)^2 \int \frac{d\omega}{2\pi}\left(2[n_\omega n_{-\omega} + (1 - n_\omega)(1 - n_{-\omega})]\left[\frac{\omega}{\mathcal{T}_k^2 + \omega^2}\right]^2 + n_\omega(1 - n_\omega)\left\{\left[\frac{2\mathcal{T}_k}{\mathcal{T}_k^2 + \omega^2}\right]^2 - \frac{8}{\mathcal{T}_k^2 + \omega^2} + \frac{4}{\mathcal{T}_k^2}\right\}\right), \quad (C5)$$

which then recovers the result of Eq. (43).

## APPENDIX D: DELTA-$T$ NOISE OF THE (1/3, 1) JUNCTION: FIRST-ORDER EXPANSION IN $\Delta T$

Rather than a numerical evaluation, it could be more rewarding to try to work out an analytic expression for the $\Delta T$ noise of the (1/3, 1) junction. This, however, can only be achieved for comparable temperatures because this approach relies on an expansion in the temperature difference $\Delta T$.

In practice, the calculation amounts to expanding the distribution function as

$$n_\omega = n_\omega^{(0)} + \frac{\Delta T}{T} n_\omega^{(1)} + O(\Delta T^2). \quad (D1)$$

Substituting this back into the expression for the noise, Eq. (45), one has

$$S(T_R, T_L) = S_0(T) + \frac{\Delta T}{T} S_1(T), \quad (D2)$$

with

$$S_0(T) = e^2 \int \frac{d\omega}{2\pi} n_\omega^{(0)} (1 - n_\omega^{(0)}) \left(1 - \frac{\mathcal{T}_k^2 - \omega^2}{\mathcal{T}_k^2 + \omega^2}\right), \tag{D3}$$

$$S_1(T) = e^2 \int \frac{d\omega}{2\pi} n_\omega^{(1)} (1 - 2n_\omega^{(0)}) \left(1 - \frac{\mathcal{T}_k^2 - \omega^2}{\mathcal{T}_k^2 + \omega^2}\right). \tag{D4}$$

The zeroth-order contribution to the distribution $n_\omega$ trivially reduces to the Fermi distribution at temperature $T$, given by

$$n_\omega^{(0)} = \frac{1}{1 + e^{\omega/T}}. \tag{D5}$$

The first-order contribution can be worked out as

$$\begin{aligned}
n_\omega^{(1)} &= \frac{3}{4} \int d\tau e^{-i\omega\tau} \frac{\omega_c}{2\pi} \frac{\sinh\left(i\pi \frac{T}{\omega_c}\right)}{\sinh\left(\pi T \left(i\frac{1}{\omega_c} - \tau\right)\right)} \left(\frac{i\pi \frac{T}{\omega_c}}{\tanh\left(i\pi \frac{T}{\omega_c}\right)} - \frac{\pi T \left(i\frac{1}{\omega_c} - \tau\right)}{\tanh\left(\pi T \left(i\frac{1}{\omega_c} - \tau\right)\right)}\right) \\
&= T \frac{3}{4} \partial_T \left[\int d\tau e^{-i\omega\tau} \frac{\omega_c}{2\pi} \frac{\sinh\left(i\pi \frac{T}{\omega_c}\right)}{\sinh\left(\pi T \left(i\frac{1}{\omega_c} - \tau\right)\right)}\right] \\
&= T \frac{3}{4} \partial_T n_\omega^{(0)}.
\end{aligned} \tag{D6}$$

It follows that

$$\begin{aligned}
n_\omega^{(1)} (1 - 2n_\omega^{(0)}) &= T \tfrac{3}{4} \partial_T n_\omega^{(0)} (1 - 2n_\omega^{(0)}) \\
&= T \tfrac{3}{4} \partial_T \left[n_\omega^{(0)} (1 - n_\omega^{(0)})\right],
\end{aligned} \tag{D7}$$

which allows us to readily write

$$S_1(T) = T \tfrac{3}{4} \partial_T S_0(T). \tag{D8}$$

Here, we have obtained an analytical expression for the delta-$T$ noise in a $(1/3, 1)$ junction. In the next Appendix we generalize this approach to a generic $(\nu_L, \nu_R)$ junction.

### APPENDIX E: UNIVERSAL FIRST-ORDER EXPANSION IN $\Delta T$

In this Appendix we report on the calculation for the expansion of $S(T_L, T_R)$ at the first other in $\Delta T/T$. The noise can be written by taking into account the temperature parametrization in Eq. (20) and up to first order in $\Delta T/T$ as

$$S(T_L, T_R) = \frac{1}{Z} \int_{-\infty}^{+\infty} d\tau \operatorname{Tr}\left\{e^{-\beta(H_L^{(0)} + H_R^{(0)})} \left[1 + \frac{\Delta T}{T^2} H_L^{(0)}\right] I(\tau) I(0)\right\} + O(\Delta T^2), \tag{E1}$$

with $\beta = T^{-1}$. Through the transformations in Eqs. (9) and (13) the noise can be rewritten as

$$S(T_L, T_R) = \frac{1}{Z} \int_{-\infty}^{+\infty} d\tau \operatorname{Tr}\left\{e^{-\beta(H_+^{(0)} + H_-^{(0)})} \left[1 + \frac{\Delta T}{T^2} H_{+,-}\right] I(\varphi_-(\tau)) I(\varphi_-(0))\right\} + O(\Delta T^2), \tag{E2}$$

with $H_{+,-}$ defined in Eq. (47) of the main text.

One needs also to consider the first-order expansion of the partition function $Z$:

$$Z = \operatorname{Tr}\left\{e^{-\beta(H_L^{(0)} + H_R^{(0)})} \left[1 + \frac{\Delta T}{T^2} H_L^{(0)}\right]\right\} + O(\Delta T^2). \tag{E3}$$

In terms of the new fields $\varphi_\pm$ and taking into account the fact that the term $(\partial_x \varphi_+)(\partial_x \varphi_-)$ gives no contribution to the trace, one has

$$Z = \operatorname{Tr}\left\{e^{-\beta H_+^{(0)}} e^{-\beta H_-^{(0)}} \left[1 + \frac{\Delta T}{2T^2} (H_+^{(0)} + H_-^{(0)} + \sin 2\theta (H_-^{(0)} - H_+^{(0)}))\right]\right\} + O(\Delta T^2), \tag{E4}$$

where we recall that $\sin 2\theta = (\nu_R - \nu_L)/(\nu_R + \nu_L)$. Then by exploiting the properties of the trace we have that

$$\begin{aligned}
Z &= Z_+^{(0)} Z_-^{(0)} + \frac{\Delta T}{2T^2}\left[-Z_-^{(0)} \frac{\partial Z_+^{(0)}}{\partial \beta} - Z_+^{(0)} \frac{\partial Z_-^{(0)}}{\partial \beta} + \sin 2\theta \left(-Z_+^{(0)} \frac{\partial Z_-^{(0)}}{\partial \beta} + Z_-^{(0)} \frac{\partial Z_+^{(0)}}{\partial \beta}\right)\right] \\
&= Z_+^{(0)} Z_-^{(0)} \left\{1 + \frac{\Delta T}{2T^2}\left[-\frac{\partial \ln(Z_+^{(0)} Z_-^{(0)})}{\partial \beta} + \sin 2\theta \frac{\partial}{\partial \beta} \ln\left(\frac{Z_+^{(0)}}{Z_-^{(0)}}\right)\right]\right\} + O(\Delta T^2),
\end{aligned} \tag{E5}$$

with

$$Z_{\pm}^{(0)} = \text{Tr}\{e^{-\beta H_{\pm}^{(0)}}\}. \tag{E6}$$

One can then define the equilibrium noise as in Eq. (E1) with $\Delta T = 0$. Since the operator $I$ only depends on $\varphi_-$ [see Eq. (5)], the trace with respect to $\varphi_+$ is trivial, leading to

$$S_0(T) = \frac{1}{Z_-^{(0)}} \int_{-\infty}^{+\infty} d\tau \, \text{Tr}\{e^{-\beta H_-^{(0)}} I(\varphi_-(\tau)) I(\varphi_-(0))\}, \tag{E7}$$

which only depends on the temperature $T$.

The expansion of the noise $S(T_L, T_R)$ up to first order in $\Delta T/T$ then reads

$$S(T_L, T_R) = S_0(T) + \frac{1}{Z_+^{(0)} Z_-^{(0)}} \int_{-\infty}^{+\infty} d\tau \, \text{Tr}\left\{ e^{-\beta(H_+^{(0)} + H_-^{(0)})} \left[ \frac{\partial \ln(Z_+^{(0)} Z_-^{(0)})}{\partial \beta} - \sin 2\theta \frac{\partial}{\partial \beta} \ln\left(\frac{Z_+^{(0)}}{Z_-^{(0)}}\right) + 2H_{+,-} \right] \right.$$
$$\left. \times I(\varphi_-(\tau)) I(\varphi_-(0)) \right\} \frac{\Delta T}{2T^2} + O(\Delta T^2). \tag{E8}$$

By exploiting the properties of the $\text{Tr}\{\ldots\}$ and after some algebra we rewrite Eq. (E8) as

$$S(T_L, T_R) = S_0(T) + \left[ S_0(T) \frac{\partial \ln(Z_+^{(0)} Z_-^{(0)})}{\partial \beta} - \sin 2\theta S_0(T) \frac{\partial}{\partial \beta} \ln\left(\frac{Z_+^{(0)}}{Z_-^{(0)}}\right) - (1 - \sin 2\theta) S_0(T) \frac{\partial \ln(Z_+^{(0)})}{\partial \beta} \right.$$
$$\left. - (1 + \sin 2\theta) \int_{-\infty}^{+\infty} d\tau \frac{1}{Z_-^{(0)}} \frac{\partial}{\partial \beta} \text{Tr}\{e^{-\beta H_-^{(0)}} I(\varphi_-(\tau)) I(\varphi_-(0))\} \right] \frac{\Delta T}{2T^2} + O(\Delta T^2). \tag{E9}$$

Since the last term can be rewritten as

$$\int_{-\infty}^{+\infty} d\tau \frac{1}{Z_-^{(0)}} \frac{\partial}{\partial \beta} \text{Tr}\{e^{-\beta H_-^{(0)}} I(\varphi_-(\tau)) I(\varphi_-(0))\} = \frac{\partial S_0}{\partial \beta} + S_0 \frac{\partial \ln Z_-^{(0)}}{\partial \beta}, \tag{E10}$$

one can finally rewrite Eq. (E8) as

$$S(T_L, T_R) = S_0(T) - (1 + \sin 2\theta) \frac{\partial S_0}{\partial \beta} \frac{\Delta T}{2T^2} + O(\Delta T^2). \tag{E11}$$

Note that from this expression we recover the noise for the particular $(1/3, 1)$ junction reported in Eq. (D8).

## APPENDIX F: RECOVERING EQ. (53)

We start from the general expression for the noise of the $(1, 1/3)$ junction, as obtained in Eq. (45), and write

$$S_0(T) = \frac{e^2}{4} \int \frac{d\omega}{2\pi} \left[ 1 - \tanh^2\left(\frac{\omega}{2T}\right) \right] \left( 2 - \frac{2\mathcal{T}_k^2}{\mathcal{T}_k^2 + \omega^2} \right), \tag{F1}$$

where we focus on the equilibrium situation, allowing us to replace the distribution function $n_\omega$ with the standard Fermi distribution at temperature $T$. We recall that the crossover energy is set by the tunneling amplitude $\Lambda'$:

$$\mathcal{T}_k = \frac{4\pi a}{v} \left(\frac{\Lambda'}{2\pi a}\right)^2. \tag{F2}$$

The first term can be readily integrated out, while the second one is reexpressed through an integration by part, thus leading to

$$S_0(T) = \frac{e^2}{\pi} T - \frac{e^2}{\pi} T \mathcal{T}_k^2 \int d\omega \tanh\left(\frac{\omega}{2T}\right) \frac{\omega}{(\mathcal{T}_k^2 + \omega^2)^2}. \tag{F3}$$

The resulting integral can then be evaluated following standard contour integration techniques, yielding

$$S_0(T) = \frac{e^2}{\pi} T - \frac{e^2 \mathcal{T}_k}{4} \frac{1}{\cos^2\left(\frac{\mathcal{T}_k}{2T}\right)} + \frac{e^2}{\pi} T \left(\frac{\mathcal{T}_k}{T}\right)^2 4\pi \sum_{n=0}^{\infty} \frac{\pi(2n+1)}{\left(\frac{\mathcal{T}_k}{T}\right)^2 - \pi^2(2n+1)^2}, \tag{F4}$$

which can further be rewritten as

$$S_0(T) = \frac{e^2}{\pi}T - \frac{e^2 \mathcal{T}_k}{4}\frac{1}{\cos^2\left(\frac{\mathcal{T}_k}{2T}\right)} + \frac{e^2}{4\pi^2}\mathcal{T}_k\left[\psi'\left(\frac{1}{2} - \frac{\mathcal{T}_k}{2\pi T}\right) - \psi'\left(\frac{1}{2} + \frac{\mathcal{T}_k}{2\pi T}\right)\right]$$

$$= \frac{e^2}{\pi}T - \frac{e^2}{2\pi^2}\mathcal{T}_k \psi'\left(\frac{1}{2} + \frac{\mathcal{T}_k}{2\pi T}\right), \tag{F5}$$

where we used the properties of the derivatives of the digamma function.

This ultimately leads back to the expression from Eq. (53) quoted in the text, namely

$$S_0(T) = \frac{1}{2}\frac{e^2}{2\pi}\mathcal{T}_k\left[4\frac{T}{\mathcal{T}_k} - \frac{2}{\pi}\psi'\left(\frac{1}{2} + \frac{\mathcal{T}_k}{2\pi T}\right)\right]. \tag{F6}$$

[1] D. C. Tsui, H. L. Stormer, and A. C. Gossard, Two-Dimensional Magnetotransport in the Extreme Quantum Limit, Phys. Rev. Lett. **48**, 1559 (1982).

[2] R. B. Laughlin, Anomalous Quantum Hall Effect: An Incompressible Quantum Fluid with Fractionally Charged Excitations, Phys. Rev. Lett. **50**, 1395 (1983).

[3] F. Wilczek, Magnetic Flux, Angular Momentum, and Statistics, Phys. Rev. Lett. **48**, 1144 (1982).

[4] D. Arovas, J. R. Schrieffer, and F. Wilczek, Fractional Statistics and the Quantum Hall Effect, Phys. Rev. Lett. **53**, 722 (1984).

[5] L. Saminadayar, D. C. Glattli, Y. Jin, and B. Etienne, Observation of the $e/3$ Fractionally Charged Laughlin Quasiparticle, Phys. Rev. Lett. **79**, 2526 (1997).

[6] R. de Picciotto, M. Reznikov, M. Heiblum, V. Umansky, G. Bunin, and D. Mahalu, Direct observation of a fractional charge, Nature (London) **389**, 162 (1997).

[7] A. M. Chang, Chiral Luttinger liquids at the fractional quantum Hall edge, Rev. Mod. Phys. **75**, 1449 (2003).

[8] M. Hashisaka, T. Ota, K. Muraki, and T. Fujisawa, Shot-Noise Evidence of Fractional Quasiparticle Creation in a Local Fractional Quantum Hall State, Phys. Rev. Lett. **114**, 056802 (2015).

[9] J. Nakamura, S. Fallahi, H. Sahasrabudhe, R. Rahman, S. Liang, G. C. Gardner, and M. J. Manfra, Aharonov-Bohm interference of fractional quantum Hall edge modes, Nat. Phys. **15**, 563 (2019).

[10] H. Bartolomei, M. Kumar, R. Bisognin, A. Marguerite, J. M. Berroir, E. Bocquillon, B. Plaçais, A. Cavanna, Q. Dong, U. Gennser, Y. Jin, and G. Fève, Fractional statistics in anyon collisions, Science **368**, 173 (2020).

[11] J. Nakamura, S. Liang, G. C. Gardner, and M. J. Manfra, Direct observation of anyonic braiding statistics at the $\nu = 1/3$ fractional quantum Hall state, Nat. Phys. **16**, 931 (2020).

[12] M. Carrega, L. Chirolli, S. Heun, and L. Sorba, Anyons in quantum Hall interferometry, Nat. Rev. Phys. **3**, 698 (2021).

[13] J. Johnson, Thermal agitation of electricity in conductors, Nature (London) **119**, 50 (1927).

[14] H. Nyquist, Thermal agitation of electric charge in conductors, Phys. Rev. **32**, 110 (1928).

[15] W. Schottky, Íber spontane stromschwankungen in verschiedenen elektrizitätsleitern, Ann. Phys. (Berlin, Ger.) **362**, 541 (1918).

[16] E. S. Tikhonov, D. V. Shovkun, D. Ercolani, F. Rossella, M. Rocci, L. Sorba, S. Roddaro, and V. S. Khrapai, Local noise in a diffusive conductor, Sci. Rep. **6**, 30621 (2016).

[17] O. S. Lumbroso, L. Simine, A. Nitzan, D. Segal, and O. Tal, Electronic noise due to temperature differences in atomic-scale junctions, Nature (London) **562**, 240 (2018).

[18] E. Sivre, H. Duprez, A. Anthore, A. Aassime, F. D. Parmentier, A. Cavanna, A. Ouerghi, U. Gennser, and F. Pierre, Electronic heat flow and thermal shot noise in quantum circuits, Nat. Commun. **10**, 5638 (2019).

[19] S. Larocque, E. Pinsolle, C. Lupien, and B. Reulet, Shot Noise of a Temperature-Biased Tunnel Junction, Phys. Rev. Lett. **125**, 106801 (2020).

[20] R. A. Melcer, B. Dutta, C. Spanslätt, J. Park, A. D. Mirlin, and V. Umansky, Absent thermal equilibration on fractional quantum hall edges over macroscopic scale, Nat. Comm., **13**, 376 (2022).

[21] A. Rosenblatt, S. Konyzheva, F. Lafont, N. Schiller, J. Park, K Snizhko, M. Heiblum, Y. Oreg, and V. Umansky, Energy Relaxation in Edge Modes in the Quantum Hall Effect, Phys. Rev. Lett. **125**, 256803 (2020).

[22] E. Zhitlukhina, M. Belogolovskii, and P. Seidel, Electronic noise generated by a temperature gradient across a hybrid normal metal-superconductor nanojunction, Appl. Nanosci. **10**, 5121 (2020).

[23] J. Rech, T. Jonckheere, B. Grémaud, and T. Martin, Negative Delta-$T$ Noise in the Fractional Quantum Hall Effect, Phys. Rev. Lett. **125**, 086801 (2020).

[24] M. Hasegawa and K. Saito, Delta-$T$ noise in the Kondo regime, Phys. Rev. B **103**, 045409 (2021).

[25] H. Duprez, F. Pierre, E. Sivre, A. Aassime, F. D. Parmentier, A. Cavanna, A. Ouerghi, U. Gennser, I. Safi, C. Mora, and A. Anthore, Dynamical Coulomb blockade under a temperature bias, Phys. Rev. Res. **3**, 023122 (2021).

[26] A. Popoff, J. Rech, T. Jonckheere, L. Raymond, B. Grémaud, S. Malherbe, and T. Martin, Scattering theory of non-equilibrium noise and delta $T$ current fluctuations through a quantum dot, J. Phys.: Condens. Matter **34**, 185301 (2022).

[27] N. Schiller, Y. Oreg, and K. Snizhko, Extracting the scaling dimension of quantum Hall quasiparticles from current correlations, Phys. Rev. B **105**, 165150 (2022).

[28] J. Eriksson, M. Acciai, L. Tesser, and J. Splettstoesser, General Bounds on Electronic Shot Noise in the Absence of Currents, Phys. Rev. Lett. **127**, 136801 (2021).

[29] G. Zhang, I. V. Gornyi, and C. Spänslätt, Delta-$T$ noise for weak tunneling in one-dimensional systems: Interactions versus quantum statistics, Phys. Rev. B **105**, 195423 (2022).

[30] B. Rosenow, I. P. Levkivskyi, and B. I. Halperin, Current Correlations from a Mesoscopic Anyon Collider, Phys. Rev. Lett. **116**, 156802 (2016).

[31] B. Lee, C. Han, and H.-S. Sim, Negative Excess Shot Noise by Anyon Braiding, Phys. Rev. Lett. **123**, 016803 (2019).

[32] M. Hashisaka, T. Jonckheere, T. Akiho, S. Sasaki, J. Rech, T. Martin, and K. Muraki, Andreev reflection of fractional quantum Hall quasiparticles, Nat. Commun. **12**, 2794 (2021).

[33] F. Guinea, Dynamics of a particle in an external potential interacting with a dissipative environment, Phys. Rev. B **32**, 7518 (1985).

[34] C. de C. Chamon, D. E. Freed, and X. G. Wen, Non-equilibrium quantum noise in chiral Luttinger liquids, Phys. Rev. B **53**, 4033 (1996).

[35] A. Schmid, Diffusion and Localization in a Dissipative Quantum System, Phys. Rev. Lett. **51**, 1506 (1983).

[36] N. P. Sandler, C. de C. Chamon, and E. Fradkin, Noise measurements and fractional charge in fractional quantum Hall liquids, Phys. Rev. B **59**, 12521 (1999).

[37] H. Ebisu, N. Schiller, and Y. Oreg, Fluctuations in Heat Current and Scaling Dimension, Phys. Rev. Lett. **128**, 215901 (2022).

[38] D. C. Tsui, Interplay of disorder and interaction in two-dimensional electron gas in intense magnetic fields, Rev. Mod. Phys. **71**, 891 (1999).

[39] X. G. Wen, Topological orders and edge excitations in fractional quantum Hall states, Adv. Phys. **44**, 405 (1995).

[40] T. Giamarchi, *Quantum Physics in One Dimension* (Oxford University Press, Oxford, 2003).

[41] E. Miranda, Introduction to bosonization, Braz. J. Phys. **33**, 3 (2003).

[42] R. Guyon, P. Devillard, T. Martin, and I. Safi, Klein factors in multiple fractional quantum Hall edge tunneling, Phys. Rev. B **65**, 153304 (2002).

[43] N. P. Sandler, C. de C. Chamon, and F. Fradkin, Andreev reflection in the fractional quantum Hall effect, Phys. Rev. B **57**, 12324 (1998).

[44] C. L. Kane and M. P. A. Fisher, Impurity scattering and transport of fractional quantum Hall edge states, Phys. Rev. B **51**, 13449 (1995).

[45] C. de C. Chamon, D. E. Freed, S. A. Kivelson, S. L. Sondhi, and X. G. Wen, Two point-contact interferometer for quantum Hall systems, Phys. Rev. B **55**, 2331 (1997).

[46] U. Weiss, R. Egger, and M. Sassetti, Low-temperature non equilibrium transport in a Luttinger liquid, Phys. Rev. B **52**, 16707 (1995).

[47] F. Guinea, G. G. Santos, M. Sassetti, and M. Ueda, Asymptotic tunneling conductance in Luttinger liquids, Europhys. Lett. **30**, 561 (1995).

[48] B. J. Overbosch and C. Chamon, Long tunneling contact as a probe of fractional quantum Hall neutral edge modes, Phys. Rev. B **80**, 035319 (2009).

[49] D. Chevallier, J. Rech, T. Jonckheere, C. Wahl, and T. Martin, Poissonian tunneling through an extended impurity in the quantum Hall effect, Phys. Rev. B **82**, 155318 (2010).

[50] A. Kamenev, *Field Theory of Non-Equilibrium Systems* (Cambridge University Press, Cambridge, 2011).

[51] C. L. Kane and M. P. A. Fisher, Thermal Transport in a Luttinger Liquid, Phys. Rev. Lett. **76**, 3192 (1996).

[52] L. Vannucci, F. Ronetti, G. Dolcetto, M. Carrega, and M. Sassetti, Interference-induced thermoelectric switching and heat rectification in quantum Hall junctions, Phys. Rev. B **92**, 075446 (2015).

[53] X. G. Wen, Edge transport properties of the fractional quantum Hall states and weak-impurity scattering of a one-dimensional charge-density wave, Phys. Rev. B **44**, 5708 (1991).

[54] P. Fendley, A. W. W. Ludwig, and H. Saleur, Exact Conductance Through Point Contacts in the $\nu = 1/3$ Fractional Quantum Hall Effect, Phys. Rev. Lett. **74**, 3005 (1995).

[55] M. P. A. Fisher and W. Zwerger, Quantum Brownian motion in a periodic potential, Phys. Rev. B **32**, 6190 (1985).

[56] C. G. Callan and M. E. Freed, Phase diagram of the dissipative Hofstadter model, Nucl. Phys. B **374**, 543 (1992).

[57] U. Weiss, Low-temperature conduction and DC current noise in a quantum wire with impurity, Solid State Comm. **100**, 281 (1996).

[58] T. Martin, *Noise in Mesoscopic Physics*, Les Houches Session LXXXI (Elsevier, Amsterdam, 2005).

[59] F. Ronetti, L. Vannucci, D. Ferraro, T. Jonckheere, J. Rech, T. Martin, and M. Sassetti, Crystallization of levitons in the fractional quantum Hall regime, Phys. Rev. B **98**, 075401 (2018).

[60] A. Kamenev and A. Levchenko, Keldysh technique and nonlinear $\sigma$-model: Basic principles and applications, Adv. Phys. **58**, 197 (2009).

[61] D. Ferraro, A. Braggio, N. Magnoli, and M. Sassetti, Neutral modes' edge state dynamics through quantum point contacts, New J. Phys. **12**, 013012 (2010).

[62] D. Ferraro, J. Rech, T. Jonckheere, and T. Martin, Single quasiparticle and electron emitter in the fractional quantum Hall regime, Phys. Rev. B **91**, 205409 (2015).

[63] S. Lal, Transport through constricted quantum Hall edge systems: Beyond the quantum point contact, Phys. Rev. B **77**, 035331 (2008).

[64] S. Roddaro, N. Paradiso, V. Pellegrini, G. Biasiol, L. Sorba, and F. Beltram, Tuning Nonlinear Charge Transport between Integer and Fractional Quantum Hall States, Phys. Rev. Lett. **103**, 016802 (2009).

[65] C. L. Kane, Matthew P. A. Fisher, and J. Polchinski, Randomness at the Edge: Theory of Quantum Hall Transport at Filling $\nu = 2/3$, Phys. Rev. Lett. **72**, 4129 (1994).

[66] D. Ferraro, M. Carrega, A. Braggio, and M. Sassetti, Multiple quasiparticle Hall spectroscopy investigated with a resonant detector, New J. Phys. **16**, 043018 (2014).

[67] N. Regnault and B. A. Bernevig, Fractional Chern Insulator, Phys. Rev. X **1**, 021014 (2011).

[68] T. Neupert, C. Chamon, T. Iadecola, L. H. Santos, and C. Mudry, Fractional (Chern and topological) insulators, Phys. Scr. **T164**, 014005 (2015).